\pdfoutput=1 
\makeatletter
\def\input@path{{tex/}}
\makeatother
\documentclass[reprint,floatfix,superscriptaddress]{revtex4-2}

\usepackage{amsmath}
\usepackage{amssymb}
\usepackage{graphicx}
\usepackage{dcolumn}
\usepackage{bm}
\usepackage{xcolor}
\usepackage{xr}

\begin{document}

\title{Nanomechanical Measurement of the Brownian Force Noise in a Viscous Liquid}

\newcommand{\BU}{Department of Mechanical Engineering, Division of Materials Science and Engineering, and the Photonics Center, Boston University, Boston, Massachusetts 02215, USA}

\author{Atakan B. Ari}
\affiliation{\BU}

\author{M. Selim Hanay}
\affiliation{Department of Mechanical Engineering, Bilkent University, 06800, Ankara, Turkey}

\author{Mark R. Paul}
\affiliation{Department of Mechanical Engineering, Virginia Tech, Blacksburg, Virginia 24061, USA}

\author{Kamil L. Ekinci}
\email[Electronic mail: ]{ekinci@bu.edu}
\affiliation{\BU}

\date{\today}

\begin{abstract}

We study the spectral properties of the thermal force giving rise to the Brownian motion of a continuous mechanical system --- namely, a nanomechanical beam resonator --- in a viscous liquid. To this end, we perform two separate sets of experiments. First, we measure the power spectral density (PSD) of the  position fluctuations of the  resonator around its fundamental mode at its center. Then, we measure the frequency-dependent linear response of the resonator, again at its center, by driving it with a harmonic force that couples well to the fundamental mode. These two measurements allow us to determine the PSD of the  Brownian force noise acting on the structure in its fundamental mode. The PSD of the force noise extracted from multiple resonators spanning a broad frequency range displays a ``colored spectrum". Using a single-mode theory, we show that, around the fundamental  resonances of the resonators, the PSD of the force noise  follows the dissipation of a blade oscillating in a viscous liquid  --- by virtue of the fluctuation-dissipation theorem.  

\end{abstract}

\maketitle

Brownian motion, the random steps taken by a micron-sized particle in a liquid, is a distinct reality of the microscopic world. The Brownian particle is incessantly bombarded by thermally-agitated liquid molecules from all sides, with the momentum exchange giving rise to a rapidly fluctuating Brownian force. One can find an approximation for the Brownian force by integrating out the many degrees of freedom of the liquid and writing a Langevin equation for  the  motion of the particle \cite{Chandler1987}. For a single-degree-of-freedom moving along the $z$ axis, the power spectral density (PSD) of the Brownian force noise $G_F(\omega)$ is related to the PSD of the particle's position fluctuations as ${G_Z}(\omega ) = {\left| {\hat \chi (\omega )} \right|^2}{G_F}(\omega )$. Here, ${ \hat \chi (\omega )}$ is the complex linear response function or the force susceptibility of the particle and describes how the particle responds to a harmonic force  at angular frequency $\omega$. The simplest description of the dynamics of the Brownian particle  comes from the assumption of a ``white" PSD for the Brownian force noise that satisfies the fluctuation-dissipation theorem \cite{Kubo}. This approximation, while neglecting all effects of inertia and flow-structure interaction, captures the long-time diffusive behavior of the Brownian particle \cite{Kubo, Franosch2011, Jannasch2011}. 

Brownian motion also sets the limits of mechanical resonators in physics experiments. Mechanical resonators with linear dimensions over many orders of magnitude --- from meter-scale mirrors~\cite{Cohadon1999, Gillespie1995, Adhikari2014} all the way down to atomic-scale nanostructures~\cite{Bunch2007, Barnard2019} -- have been used for detecting charge~\cite{Cleland1998} and mass~\cite{Ekinci2004a}, and for studying electromagnetic fields \cite{Aspelmeyer2014} and  quantum mechanics~\cite{OConnell2010}. A typical continuous mechanical resonator can be described  as a collection of mechanical modes, with each mode behaving like a particle bound in a harmonic potential, i.e., a harmonic oscillator~\cite{Cleland2003}. With the normal mode approximation, Brownian motion of a continuous mechanical resonator can be easily formulated  for small dissipation~\cite{Saulson1990, Cleland2002}, when spectral flatness (i.e., a white PSD)  and modal orthogonality  can be assumed for the Brownian force. For a multi-degree-of-freedom system with large and spatially varying dissipation, however, the theoretical formulation of Brownian motion is non-trivial~\cite{Majorana1997, Gillespie1995}. Here, the modes are coupled strongly and motions in different modes become correlated~\cite{Saulson1990,Schwarz2016}. A possibility for finding the characteristics of the thermal force comes from the fluctuation-dissipation theorem, assuming that one can determine the dissipation in the system from separate theory, e.g., fluid dynamics~\cite{Paul2004}. 

While the Brownian force  acting on single-degree-of-freedom particles has  been  directly measured  in liquids \cite{Franosch2011, Jannasch2011, Mo:2015}, the few reports on  continuous mechanical resonators remain in the small dissipation limit~\cite{Miao:2012, Teufel:2009, Doolin:2014}. In the  presence of large dissipation, experimental challenges, such as dampened signal levels and lack of reliable motion actuation methods, have so far precluded the direct measurement of the Brownian force on continuous mechanical systems.  Here, we employ optical and electronic measurement techniques to  extract the PSD of the Brownian force noise acting on a nanomechanical  resonator  in a viscous liquid.  The force noise exerted by the liquid on the resonator has a ``colored" PSD and follows the viscous dissipation of the resonator as dictated by the fluctuation-dissipation theorem~\cite{Paul2004}. A single-mode approximation obtained from fluid dynamics \cite{Cross2006} captures the observed colored PSD at low frequency but deviates from the experiment with increasing frequency where higher mode contributions and details of our external driving approach become appreciable.

\begin{figure*}
   \includegraphics[width=6.75in]{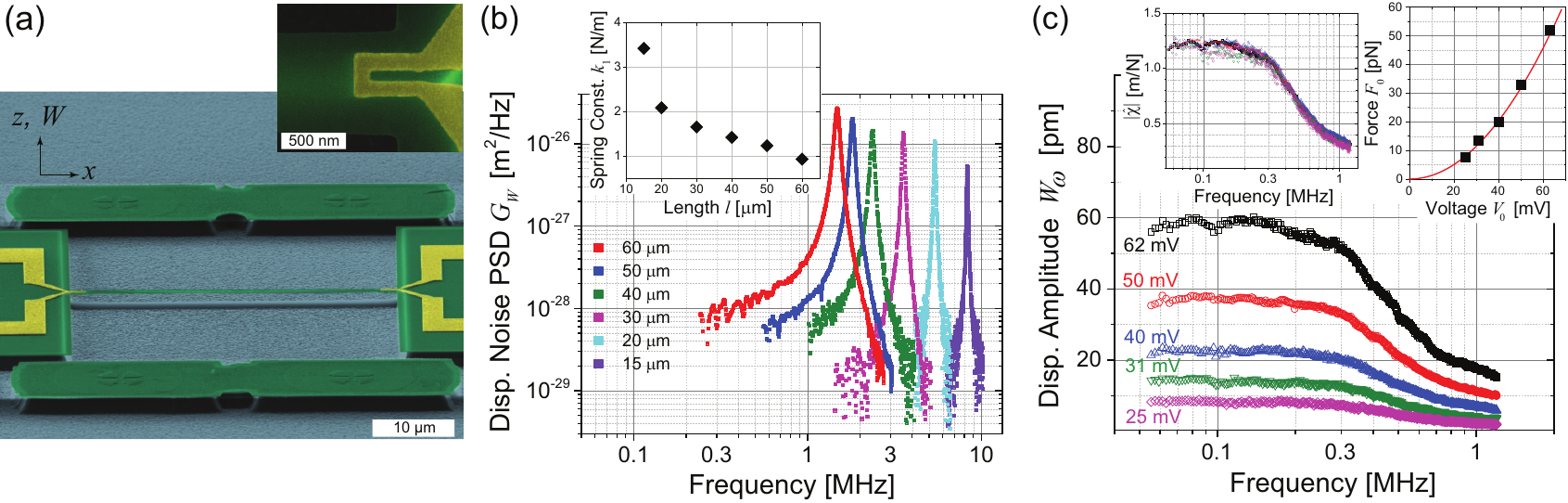} 
    \caption{(a)~False-colored SEM image of a silicon nitride nanomechanical beam ($l \times b \times h \approx 40~\mu \rm m \times 950~ \rm nm \times 93~ \rm nm$), showing the structure (green) and the actuators (yellow).  (Inset) Close-up image of an electrothermal actuator near one of the clamps of the beam. This is a u-shaped gold film resistor deposited on top of the beam with a thickness of $ 100~\rm nm$ and a width of $120~\rm nm$. (b)~PSDs of the displacement noise of beams with different lengths ($15~{\rm\mu m} \le l \le 60~{\rm\mu m}$) measured in air around their fundamental resonant modes and at their centers ($x=l/2$). The inset shows the spring constant $k_1$ as a function of length for each beam. (c)~Driven displacement amplitudes of the 60-$\rm \mu m$-long beam in water at different drive voltages $V_0$. Left inset shows the magnitude of the susceptibility $ \left|{\hat \chi} \right|$ as a function of frequency, and the right inset shows the responsivity of the force transducer. The line is a fit to $F_0 = A {V_0}^2$ with $A=1.27 \times 10^{-8}~\rm N/V^2$.}
    \label{fig:Figure1}
\end{figure*}

Our experiments are performed on nanomechanical silicon nitride  doubly-clamped beam resonators under tension. Figure~\ref{fig:Figure1}(a) shows a false-colored scanning electron microscope (SEM) image of a typical beam that has  dimensions of $l \times b \times h \approx  \rm 40~\mu m \times950~nm \times 93~nm$. There is a  2 $\rm \mu m$ gap between the beam and the substrate. There are two  u-shaped metal (gold)  electrodes on each end of the beam, through which an AC electric current can be passed (Fig.~\ref{fig:Figure1}(a) inset). This causes Ohmic heating cycles, which in turn generate  thermal bending moments. The result is efficient actuation of nanomechanical oscillations at exactly twice the frequency of the applied AC current \cite{Ari2018, Bargatin2007a}. Both the driven and Brownian motions of the beams are measured using a path-stabilized Michelson interferometer that can resolve a displacement of $\sim 5~ \rm fm/ \sqrt{Hz}$ after proper numerical background subtraction \cite{Kara2015}.  We focus on the out-of-plane motions of the beams  at their centers ($x = l/2$), denoted by ${W}(t)$ (Fig.~\ref{fig:Figure1}(a)). For measurements in liquid, the device chip is immersed in a small bath of liquid. Table~\ref{tab_devices} lists the dimensions and experimentally-determined mechanical parameters of all the devices used in this study.  In the following analysis, we use a density of $\rho_s=2750$ kg/m$^3$, Young's modulus of $E=300$ GPa, and a tension force of $S=7.64$ $\mu$N for all the beams. All experimental details and data are available in the SI~\cite{Supp}. 

\begin{table}
\caption{\label{tab_devices} Experimentally-obtained mechanical properties of the measured devices.}
\begin{ruledtabular}
\begin{tabular}{ccccccc}
Device &   $l \times b \times h$ & $k_1$  & $m_1$ & $\omega_1 / 2\pi$ \\
     & ($\mu \rm m^3$)       & (N/m) &    (pg)  & (MHz)   \\
\hline
$60 \mu \rm m$ &  $60 \times 0.95 \times 0.093$ &  0.93  & 10.82  & 1.48  \\
$50 \mu \rm m$ &  $50 \times 0.95 \times 0.093$ &  1.24 & 9.63  & 1.80  \\
$40 \mu \rm m$ &   $40 \times 0.95\times 0.093$ &  1.42 & 6.53  & 2.35 \\
$30 \mu \rm m$ &   $30 \times 0.95\times 0.093$ &  1.66 & 3.33 & 3.55 \\
$20 \mu \rm m$ &   $20 \times 0.95\times 0.093$ &  2.09 & 1.83  & 5.38 \\
$15 \mu \rm m$ &   $15 \times 0.95\times 0.093$ &  3.42 & 1.26  & 8.28 \\
\end{tabular}
\end{ruledtabular}
\end{table}

We first describe how the spring constants $k_1$ and effective masses $m_1$ are obtained for the fundamental mode of the resonators from thermal noise measurements in air. Figure~\ref{fig:Figure1}(b) shows the PSD of the displacement noise (position fluctuations) $G_{W}$ (in units of $\rm m^2/Hz$) at $x=l/2$ as a function of frequency $\frac{\omega}{2 \pi}$ for each resonator around its fundamental mode resonance frequency. Since the resonances are sharply-peaked, $G_{W}$ can be integrated accurately over frequency to obtain the  mean-squared fluctuation amplitude, $\langle {W}^2 \rangle$, for the fundamental mode. Using the classical equipartition theorem, we determine the spring constants of the resonators as $k_1  = k_B T /\langle {W}^2 \rangle$, where $k_B$ is the Boltzmann constant and $T$ is the temperature. Thus, $k_1$ is the spring constant for the fundamental mode when measured at $x=l/2$. The inset of Fig.~\ref{fig:Figure1}(b) shows $k_1$ as a function of beam length. In air, the frequency of the fundamental mode $\omega_1$ and its effective mass $m_1$ are assumed to be very close to their respective values in vacuum~\cite{Kara}. Thus, with $k_1$ and $\omega_1$ in hand, $m_1$ can be found from $m_1=k_1/ {\omega_1}^2$ where $m_1$ is the effective mass of the beam in the absence of a surrounding liquid. We discuss how the experimentally measured values of $k_1$, $m_1$ and $\omega_1$ relate to the theoretical predictions for an Euler-Bernoulli beam under tension in the SI~\cite{Supp}. 

\begin{figure*}
\includegraphics[width=6.75in]{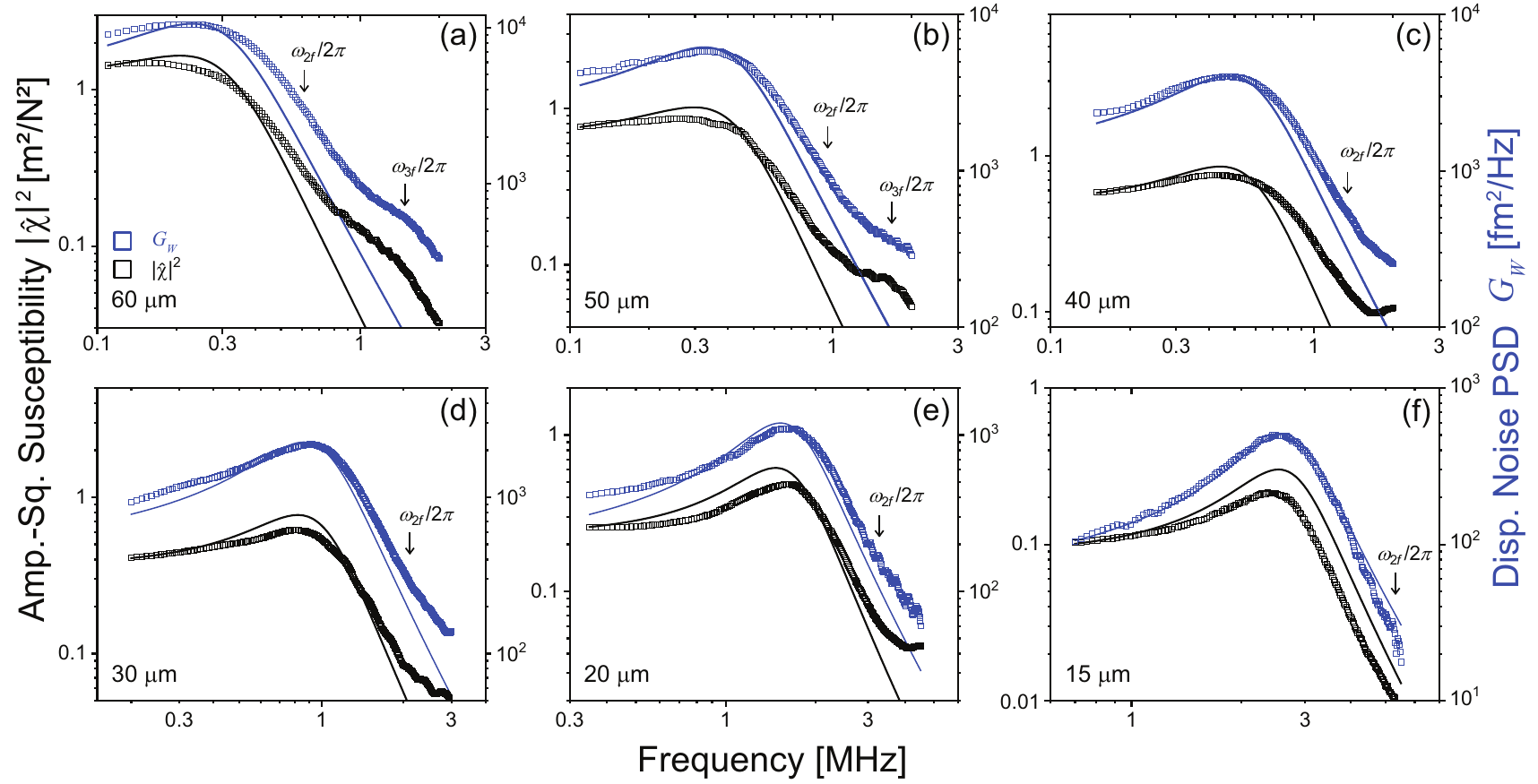} 
    \caption{Amplitude-squared susceptibility $\left|{\hat \chi} \right|^2$ (black) and PSD of the displacement fluctuations $G_{W}$ (blue) for each beam in water; beam length is indicated in the lower left of each panel. Both quantities are measured at the center of the beam $x=l/2$. Square symbols are experimental measurements.  The continuous lines are theoretical predictions using the single-mode description.  The arrows show the approximate positions of the peaks of the second ($\omega_{2f}/2\pi$) and third mode ($\omega_{3f}/2\pi$) in fluid (when in range).}
    \label{fig:Figure2}
\end{figure*}

We now turn to the calibration of the forced response in water. Under a harmonic force $F(t) = {F_0}\sin {\omega t}$, we can write the oscillatory displacement of the beam at its center as ${W}(t) = {{W}_\omega }\sin \left( {\omega t + {\varphi_\omega }} \right)$, with ${W}_\omega$ and $\varphi_\omega$ being the frequency-dependent displacement amplitude and phase, respectively. Linear response theory yields ${{W}_\omega } = \left| {\hat \chi (\omega )} \right|{F_0 }$ \cite{Sethna2006}.  Assuming that the fundamental mode response dominates at low frequency ($\omega \to 0$), one recovers the familiar static (DC) response ${{W}_{dc}} = {F_0}/{k_1}$ (see Eqs.~(\ref{Linear-response-general}) and~(\ref{chi_d}) below). Figure~\ref{fig:Figure1}(c) shows the driven response of the  60-$\rm\mu m$ beam at $x=l/2$ in water obtained at  several different drive voltages. Each data trace is collected by applying to the electrothermal actuator a sinusoidal voltage with constant amplitude $V_0$ and sweeping the frequency of the voltage. The displacement amplitude at low frequency, ${W}_{dc}$, is determined from each trace, and $F_0$ is found as $F_0\approx k_1 {{W}_{dc}}$ with $k_1$ from thermal noise measurements. From the measured displacement amplitudes at different drive amplitudes, one can  obtain the force transducer responsivity, shown in the upper right inset of Fig.~\ref{fig:Figure1}(c).  More importantly, under the assumption that $F_0$ is constant~\cite{Supp}, one can extract the amplitude of the force susceptibility as $\left|{\hat \chi (\omega )} \right|\approx {W}_\omega/F_0$. The  left inset of Fig.~\ref{fig:Figure1}(c) shows that the  response of the device at different drive voltages (forces) can be collapsed  onto  $\left|{\hat \chi (\omega )} \right|$ as measured at its center using the proper force calibration. 

We show measurements of the driven response and the Brownian fluctuations  for each nanomechanical beam  in water in Fig.~\ref{fig:Figure2}. Each double-logarithmic plot corresponds to a separate beam, showing the PSD of the displacement noise, $G_{W}$ (blue,  $\rm fm^2/Hz$), and the amplitude-squared susceptibility,  ${\left| \hat \chi \right|^2}$ (black, $\rm m^2/N^2$), as a function of frequency. In all plots, the range shown for both $G_{W}$ and ${\left| \hat \chi \right|^2}$ are adjusted to span exactly two decades; the ranges of frequency shown are different. The approximate positions of the peaks of the second mode (not detectable at the center of the beam) and the third mode are marked with arrows when in range. Several important observations can be made. It is evident that $G_{W}$ and ${\left| \hat \chi \right|^2}$ show different frequency dependencies and peak positions. It is precisely these variations that we will use to provide an experimental estimate of the PSD of the Brownian force noise. It is also clear from Fig.~\ref{fig:Figure2} that, as the frequency of the fundamental mode increases for the different devices of decreasing length, the overdamped response progressively turns underdamped  at higher frequencies where the mass loading due to the fluid is reduced~\cite{Cross2006}.  
\begin{figure*}
   \includegraphics[width=6.75in]{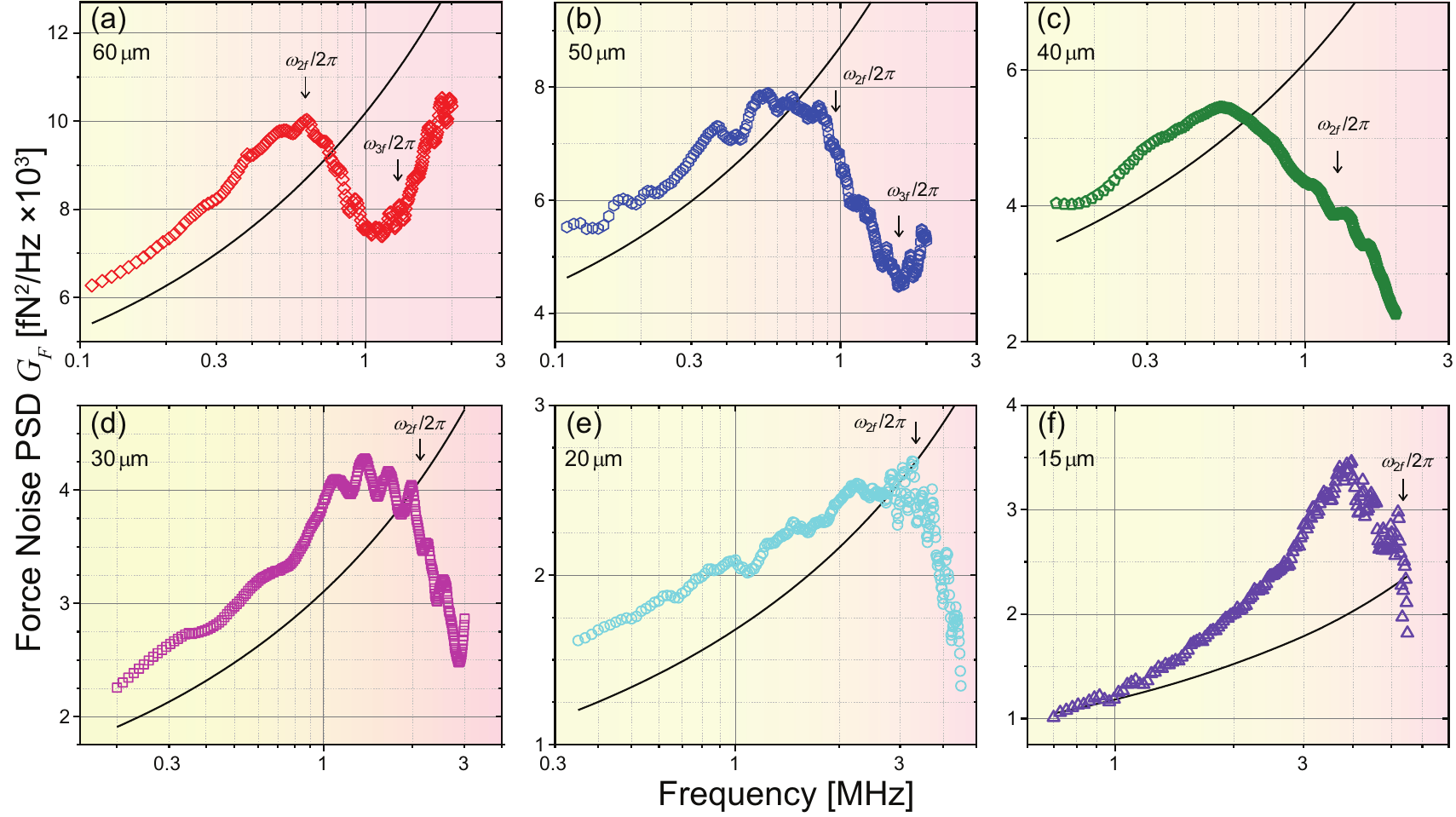} 
    \caption{PSD of the Brownian force $G_F$ acting on each beam in water. The  line is the prediction of  Eq.~(\ref{GF}) for an oscillating blade in a viscous fluid. Arrows show the higher mode peak positions as in Fig.~\ref{fig:Figure2}. The shading indicates the approximate frequency region where the fundamental mode is dominant. }
    \label{fig:Figure3}
\end{figure*}

The continuous lines in Fig.~\ref{fig:Figure2} are from a theoretical description that treats the beams as single-degree-of-freedom harmonic oscillators in a viscous fluid~\cite{Cross2006}. We first express the linear response function in the familiar general form
\begin{equation}
    \label{Linear-response-general}
	{\left| {\hat \chi (\omega )} \right|^2} = \frac{1}{{{{\left[ {{k_1} - {m_f}(\omega ){\omega ^2}} \right]}^2} + {\omega ^2}{{\left[ {{\gamma _f}(\omega )} \right]}^2}}}.
\end{equation}
The modal mass $m_f$ in fluid is a function of frequency due to the mass of fluid that is moving in conjunction with $m_1$. In addition, the dissipation due to the viscous fluid $\gamma_f$ is frequency dependent.  Both $m_f(\omega)$ and $\gamma_f(\omega)$ can be determined from fluid dynamics by approximating the beam as a long and slender blade (or cylinder) oscillating perpendicular to its axis in a manner consistent with the fundamental mode amplitude profile of the beam~\cite{Rosenhead,sader:1998,Cross2006}. This description yields $\gamma_f(\omega) = m_1 T_0 \omega \Gamma_b''({\rm Re}_\omega)$ and ${m_f}(\omega) = {m_1}\left( {1 + {T_0}\Gamma_b'({\rm{R}}{{\rm{e}}_\omega })} \right)$. The hydrodynamic function of the blade $\Gamma_b$ is expressed as a function of the frequency-based Reynolds number, ${\rm Re}_\omega=\frac{\omega b^2}{4 \nu_f}$, where $\nu_f$ is the kinematic viscosity of the fluid~\cite{sader:1998,Cross2006}.  $\Gamma_b$ is a complex valued function, $\Gamma_b({\rm Re}_\omega) = \Gamma_b'({\rm Re}_\omega) + i\Gamma_b''({\rm Re}_\omega)$, that is determined as an ${\cal O} (1)$ correction to the hydrodynamic function of an oscillating cylinder in fluid~\cite{sader:1998,Clark2010}.  The mass loading parameter, $T_0 = \frac{\pi \rho_f b}{4 \rho_s h}$, is the ratio of the mass of a cylinder of fluid with diameter $b$ to the mass of the beam, where $\rho_f$ and $\rho_s$ are fluid and solid densities, respectively. Using these ideas, the amplitude-squared susceptibility can be expressed as~\cite{Cross2006} 
\begin{equation} \label{chi_d}
{\left| {\hat\chi (\omega )} \right|^2} = \frac{1}{{{{\left[ {{k_1} - {m_1}\left( {1 + {T_0}\Gamma_b'} \right){\omega ^2}} \right]}^2} + {{{\omega ^2} \left[ {{m_1}{T_0}\omega \Gamma_b''} \right]}^2}}}
\end{equation}
for the fundamental mode of the beam when measured at $x=l/2$. Comparing Eq.~(2) with Eq.~(1), one can clearly see how the oscillating blade solution provides the parameters for the single-degree-of-freedom  harmonic oscillator.

The PSD of the position fluctuations of the fundamental mode of the beam can be expressed as
\begin{equation}
   G_{{W}}(\omega)=\left| {\hat \chi (\omega )} \right|^2 G_F(\omega). 
   \label{Eq:Disp_Noise_PSD}
\end{equation}
It follows from the fluctuation-dissipation theorem~\cite{callen:1951,callen:1952} that the PSD of the  Brownian force noise for an oscillating blade or cylinder in fluid can be expressed as~\cite{Paul2004}
\begin{equation}
    G_F(\omega) = 4 k_B T  m_1 T_0 \omega \Gamma_b''({\rm Re}_\omega).
\label{GF}
\end{equation}
In Fig.~\ref{fig:Figure2}, the black lines use Eq.~(\ref{chi_d}) and the blue lines use Eqs.~(\ref{Eq:Disp_Noise_PSD}-\ref{GF}) where $k_1$ and $m_1$ are measured from experiment; the force is calibrated using $k_1$ at zero frequency; and $\Gamma_b$ and $T_0$ are calculated from the dimensions and density of the beam and the properties of water. In other words, there are \textit{no} free fit parameters.
\begin{figure}
   \begin{center}
   \includegraphics[width=3.375in]{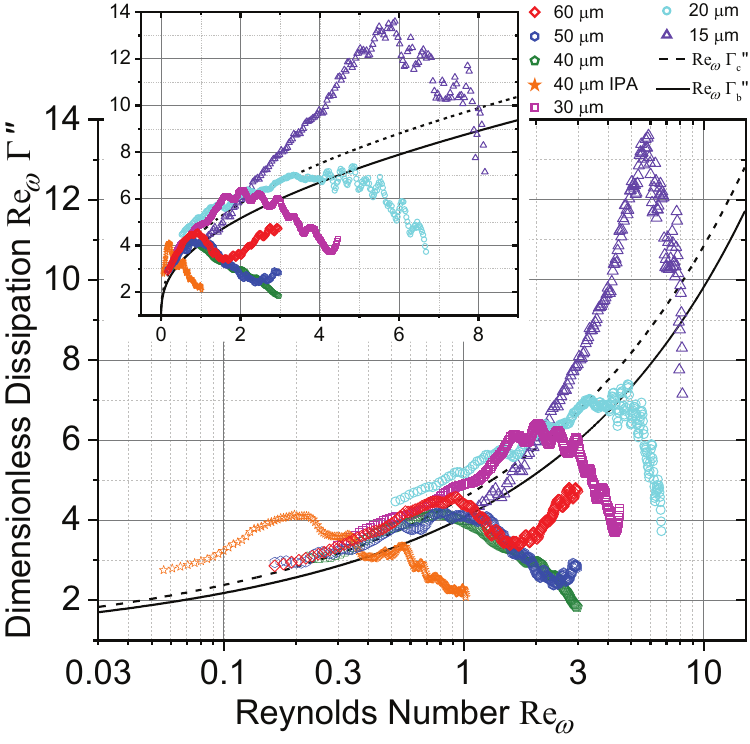} 
    \caption{ Semi-logarithmic and linear (inset) plots showing all the data in the dimensionless dissipation form. Theoretical predictions are shown for an oscillating cylinder (dashed line) and for an oscillating blade (solid line).}
    \label{fig:Figure4}
    \end{center}
\end{figure}

With the displacement noise PSD and the susceptibility experimentally determined, we can find an estimate of the PSD of the Brownian force noise exerted on the beams by the surrounding liquid using $G_F(\omega)=G_{W}(\omega)/ |\hat \chi(\omega)|^2$. The symbols in Fig.~\ref{fig:Figure3}(a)-(f) show the experimentally-obtained PSDs of the Brownian force (in units of $\rm fN^2/Hz$) in water.  The monotonically-increasing continuous lines are the theoretical predictions for an oscillating blade in a viscous fluid given by Eq.~(\ref{GF}). The arrows indicate the approximate peak frequencies, $\omega_{2f}$ and $\omega_{3f}$, of the higher modes as in Fig.~\ref{fig:Figure2}. The experimental data in Fig.~\ref{fig:Figure3}(a)-(f) increase with frequency for $\omega \lesssim \omega_{2f}$ as predicted by the theory of an oscillating blade in fluid.  However, the experiment begins to deviate from theory around $\omega_{2f}$ for all resonators (Fig.~\ref{fig:Figure3}(a)-(f)). After making a dip, the data in Fig.~\ref{fig:Figure3}(a) and (b) begin to increase again around $\omega_{3f}$; this feature remains out of the measurement range in Fig.~\ref{fig:Figure3}(c)-(f). We attribute these deviations from the theoretical prediction to the influence of the higher modes of oscillation in the driven response of the beam~\cite{Supp}, in qualitative agreement with~\cite{Clark2010}.

The general trends of the Brownian force can be made clearer by plotting the experimental data nondimensionally. From Eq.~(\ref{GF}), the dimensionless variation of the force PSD can be expressed as ${\rm Re}_\omega \Gamma''({\rm Re}_\omega)$.  Figure~\ref{fig:Figure4} shows ${\rm Re}_\omega \Gamma''({\rm Re}_\omega)$ for each data trace in Fig.~\ref{fig:Figure3} using semi-logarithmic (main) and linear (inset) plots. To make this plot, we have found  experimental  $\Gamma''$ values   from $\Gamma'' = G_F/(4k_B T m_1 T_0 \omega$)  for each beam, with ${\rm Re}_\omega$ acting as a nondimensional frequency.  Also shown in Fig.~\ref{fig:Figure4} are the theoretical predictions for an oscillating cylinder (dashed line) and blade (solid line) in a viscous fluid.  In addition to the six data sets in water, we include data taken in isopropyl alcohol (IPA)~\cite{Supp}. IPA, with its higher viscosity and lower density compared to water, allows us to extend our dimensionless parameter space.  The data in Fig.~\ref{fig:Figure4} extend over two decades in ${\rm Re}_\omega$ and follow the viscous dissipation of a blade (or cylinder) oscillating in the liquid over a range of frequencies. 

A more accurate description of the Brownian dynamics of the continuous beams used in the experiments should include the contributions from the higher modes. If we use a multimode lumped description, we can express the PSD of the displacement fluctuations  as   (cf. Eq.~(\ref{Eq:Disp_Noise_PSD}))
\begin{equation}
G_{{W}}(\omega) = \sum_{n=1,3,\ldots}^{\infty} |\hat{\chi}_n(\omega)|^2 G_{F,n}(\omega),
\label{eq:gwsum}
\end{equation}
where $\hat \chi_n(\omega)$ represents  the susceptibility of the $n^{\rm th}$ mode and the even modes do not contribute since the measurement is at $x=l/2$. Eq.~(\ref{eq:gwsum}) is the sum of the different modal contributions where the modes are assumed to be uncorrelated with one another. For small mode number $n$, a reasonable assumption is that $G_{F,n}(\omega)$ is independent of $n$ to yield $G_{{W}}(\omega) \approx G_{F}(\omega) \sum |\hat{\chi}_n(\omega)|^2$, where it is clear that the  important quantity is the sum of the squares of the susceptibilities of the individual modes. 

Similarly, the driven response should be described using a multimode approach that accounts for the spatially-varying aspects of electrothermal drive. This analysis  yields an approximate expression of the form
\begin{equation}
{{W}_\omega}^2 \approx \bar{\alpha} \left( \pi F_0 \right)^2 \left|\sum_n \psi_n \hat{\chi}_n(\omega) \right|^2.
\label{eq:wsum}
\end{equation}
Here, $F_0$ is the magnitude of the electrothermal  force and $\psi_n$ accounts for the coupling of the drive to mode $n$ \cite{Supp}. We have used the fact that  magnitude of the odd mode shapes at the center of the beam are nearly constant in order to factor out the constant $\bar{\alpha} \approx \left( \phi_n(1/2) \right) ^{-2}$ for small and odd $n$, where $\phi_n(1/2)$ is the normalized mode shape evaluated at the center of the beam.

Equations~(\ref{eq:gwsum})-(\ref{eq:wsum}) provide some insight into the deviations of the experiment from  theory observed in Figs.~\ref{fig:Figure3} and~\ref{fig:Figure4}. For the electrothermal drive applied at the distal ends of the beam, the coefficients $\psi_n$ are not expected to be constant and  will result in non-trivial contributions from the higher modes. Since, in the single mode approximation, we estimate $G_F (\omega)$ by dividing the displacement noise PSD at $\omega$ by the driven response at the same $\omega$,  variations in the driven response due to $\psi_n$  result in deviations from expected behavior for high frequencies where the influence of the higher modes are significant. We point out that the effects of $\psi_n$ cannot be simply deconvoluted or factored out. The electrothermal driven response of the higher modes are entangled due to the large damping in the system, and the driven response becomes quite complicated as the frequency increases.  We highlight that Eq.~(\ref{eq:wsum}) contains the square of the sum whereas Eq.~(\ref{eq:gwsum}) is the sum of the squares. As a result, Eq.~(\ref{eq:wsum}) would contain complicated contributions due to the cross terms even if the coefficients $\psi_n$ could be made nearly constant.
  
 This first direct measurement of the PSD of the Brownian force in a liquid  employing nanomechanical resonators is a remarkable manifestation of the fluctuation-dissipation theorem. Even a qualitative explanation of the experimental deviation from theory has required consideration of subtle aspects of the driven response of a continuous system.  In the near future, a transducer capable of exerting forces at arbitrary positions with high spatial resolution~\cite{Sampathkumar2006} may allow for directly determining $|\hat \chi_n (\omega)|^2$ for several individual modes.  This could then be used to extend the frequency range of the type of measurements described here and would lead to further physical insights into the Brownian force acting on a continuous nanostructure.

ABA and KLE acknowledge support from US NSF through Grant Nos. CBET-1604075 and CMMI-2001403. MRP acknowledges support from US NSF Grant No. CMMI-2001559. 
\bibliography{main.bib}

\end{document}


\title{Supplementary Material for\\ ``Nanomechanical Measurement of the Brownian Force Noise in a Viscous Liquid''}

\newcommand{\BU}{Department of Mechanical Engineering, Division of Materials Science and Engineering, and the Photonics Center, Boston University, Boston, Massachusetts 02215, USA}

\author{Atakan B. Ari}
\affiliation{\BU}

\author{M. Selim Hanay}
\affiliation{Department of Mechanical Engineering, Bilkent University, 06800, Ankara, Turkey}

\author{Mark R. Paul}
\affiliation{Department of Mechanical Engineering, Virginia Tech, Blacksburg, Virginia 24061, USA}

\author{Kamil L. Ekinci}
\email[Electronic mail: ]{ekinci@bu.edu}
\affiliation{\BU}

\maketitle

\tableofcontents

\newpage

\section{Fabrication and Device Properties}

\subsection{Material Properties}

Our resonators are fabricated out of a 100-$\rm nm$ silicon nitride film that is deposited on top a 550-$\mu \rm m$-thick silicon handle wafer~\cite{Wafer} via low-pressure chemical vapor deposition (LPCVD).  The  material properties of silicon nitride depend upon the specifics of the LPCVD process. The reported Young's modulus and density values are in the range  $250~{\rm GPa} \le E \le 350 ~ {\rm GPa}$ and $2600~ {\rm kg / m^3} \le \rho_s \le 3400 ~ {\rm kg / m^3}$, respectively \cite{Senturia:2002, Drummond:1995, Santucci:2001, Stoffel:1996, Larsen:2013, Fauver:1998, Petersen:1982, Edwards:2004, Yang:2008, Yang:2002, Tabata:1989}.  Furthermore, there is an unknown stress in the silicon nitride film, which results in a tension force $S$ in the suspended beams. 

In Section \ref{section:Determining Material Properties from Natural Frequencies}, we show that beam theory and the measured eigenfrequencies of the beams allow us to estimate a single set of optimal material properties that can be used to describe all of the devices used in our experiments. We have found that this approach is more insightful than trying to estimate the specific material properties of each beam individually. The values we have found from this analysis are $E=300~ \rm GPa$, $\rho_s=2750~ \rm kg/m^3$,  $\sigma = 86.5~ \rm MPa$, and $S= 7.64~ \rm  \mu N$, as listed in Tables~\ref{tab_devices} and \ref{fit_parameters}. 

\subsection{Fabrication Process}

\begin{figure}[b]
   \begin{center}
   \includegraphics[width=6.75in]{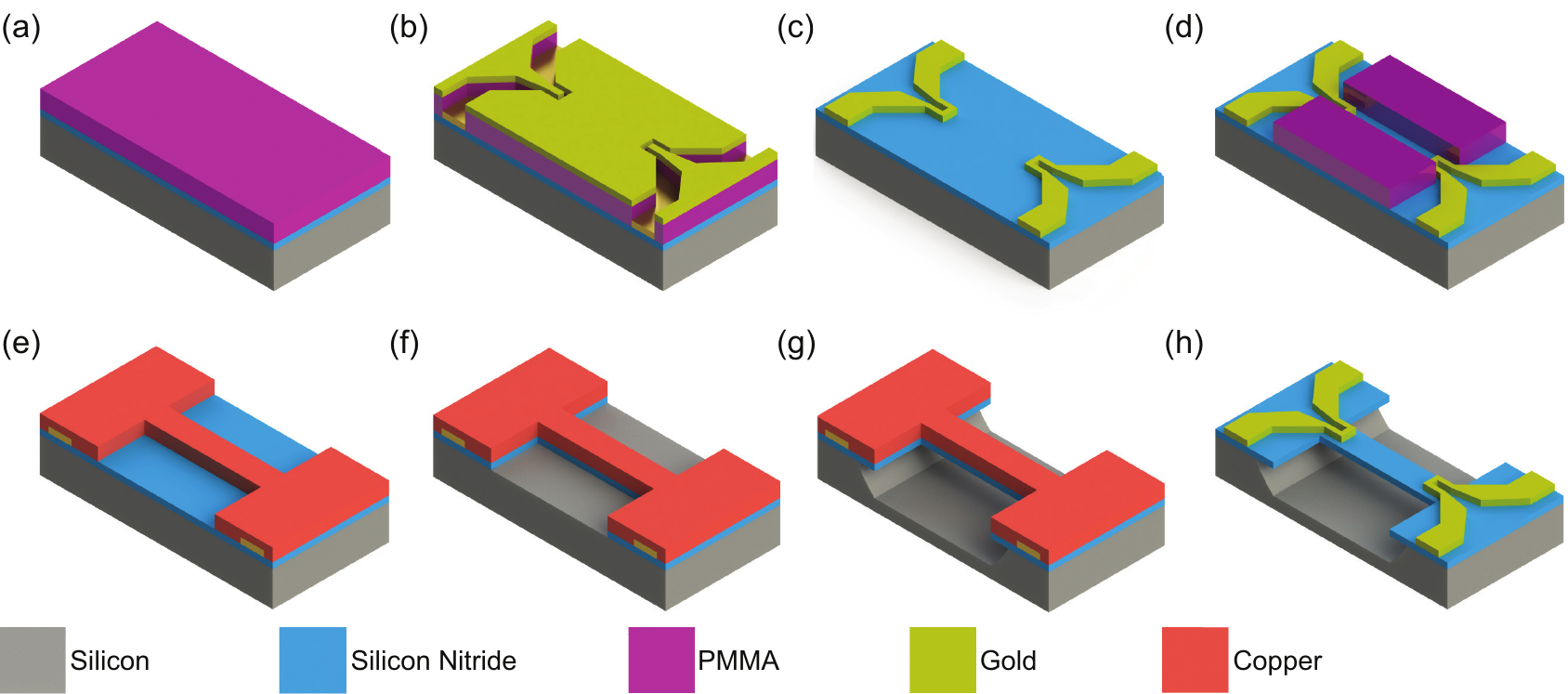} 
    \caption{Flowchart for the fabrication process. (a) PMMA resist is spun on the silicon nitride layer, (b) then transducers are patterned using EBL, followed by deposition of gold. (c) After lift-off, the transducers and contact pads are completed. (d) A second step of EBL is used to pattern the nanomechanical structures. (e) To mask the beam structures during dry etch, a copper etch mask is deposited. (f) Silicon nitride is etched anisotropically down towards the silicon layer, and then (g) silicon is etched isotropically to suspend the beams. (h) After the etch mask is removed, the fabrication is complete.}
    \label{fig:Supp_fabrication}
    \end{center}
\end{figure}

\begin{figure}
   \begin{center}
   \includegraphics[width=6.75in]{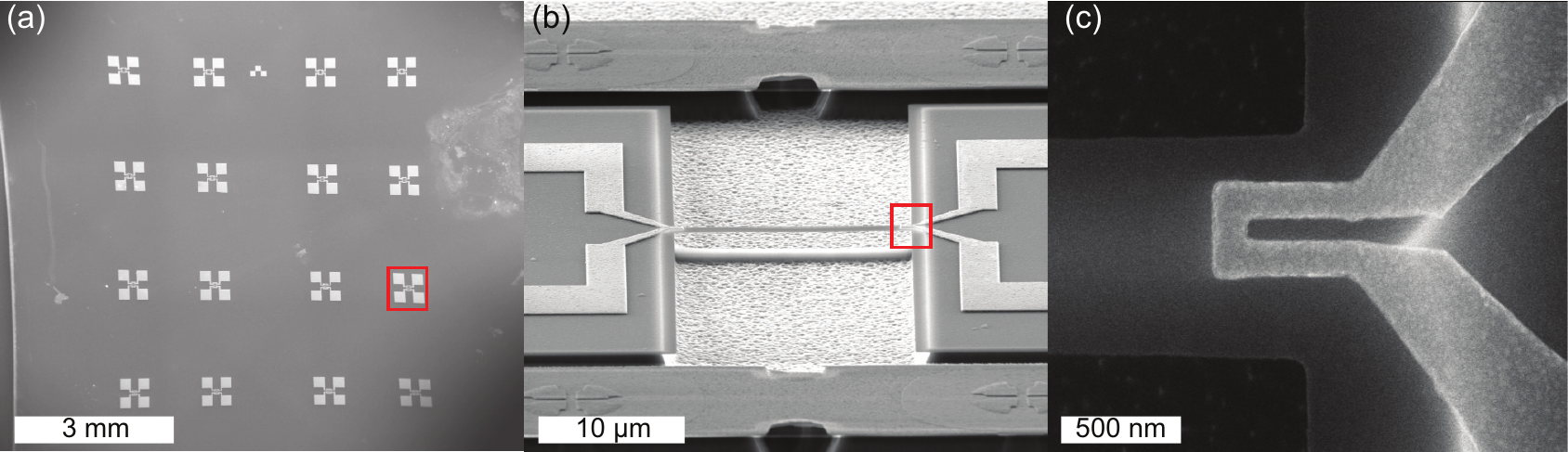} 
    \caption{(a) A $10~\rm mm \times 10 ~\rm mm$ chip that has 16 NEMS devices with lengths 15~${\rm\mu m} \le l \le$ 60~${\rm\mu m}$. (b) Close-up and tilted SEM image of a 15-$\rm \mu m$-long beam. The gap between the suspended beam and the substrate is approximately 2 $\rm \mu m $. (c) Magnified top view of a transducer.}
    \label{fig:Supp_SEM}
    \end{center}
\end{figure}

A flowchart of the fabrication process is given in Fig.~\ref{fig:Supp_fabrication}. Transducers and contact pads are patterned using electron beam lithography (EBL). A 100-nm-thick gold film with a 5-nm chromium adhesion layer underneath is deposited by thermal evaporation. After lift-off, a second step of EBL is performed to define the beam structures. A 60-nm-thick copper film is deposited via electron beam evaporation, which serves as a dry etch mask. Inductively coupled plasma  etching is used to etch the silicon nitride layer anisotropically; then, the beam is suspended by an isotropic etch step that removes the silicon beneath. After the removal of the copper etch mask, the devices are placed in a chip carrier with 50-$\Omega$ strip lines and wirebonds are made between the contact pads and the striplines. Figure~\ref{fig:Supp_SEM} shows SEM images of a completed chip  and one of the resonators, a 15-$\rm \mu m$-long beam. 

\subsection{Device Properties}

In this section, we provide more details on the device dimensions and mechanical properties and parameters. The dimensions of the beams reported in Table~\ref{tab_devices} are determined by imaging them in SEM at high magnification. The in-plane dimensions are found to be close to the design dimensions with small deviations due to fabrication imperfections. The  thickness of the silicon nitride beams has more uncertainty and  is measured to be $\rm 92.8 \pm 0.8~nm$. In the theoretical calculations, we have used  the value $\rm 92.8~nm$.  In order to find the mechanical parameters, we measure the first few resonant frequencies by driving the resonators in air; we also  measure the thermal vibrations of the fundamental mode of each resonator in air and find the modal masses and spring constants by the equipartition theorem. We observe a close agreement between the mean-squared fluctuation amplitudes in air and water, as  expected from the equipartition theorem. This is a good sanity check and proves that the spring constant does not change in water. All the relevant device and material parameters in this work can be found in Tables \ref{tab_devices}, \ref{tab_device_frequencies} and \ref{fit_parameters}.
\begin{table}[b!]
\caption{\label{tab_devices} Device parameters and experimentally-obtained mechanical properties of the  devices used in this study. The mean-squared fluctuation amplitudes measured at the centers of the beams in air and water, $\left\langle {{{W}_{air}}^2} \right\rangle$ and $\left\langle {{{W}_{water}}^2} \right\rangle $, respectively, are listed for comparison.}
\begin{ruledtabular}
\begin{tabular}{cccccccccc}
Device &   $l \times b \times h$ & $k_1$  & $m_1$  & $\alpha_1$ & $\omega_1 / 2\pi$ & $U$ & $S$ & $\left\langle {{{W}_{air}}^2} \right\rangle $ & $\left\langle {{{W}_{water}}^2} \right\rangle $ \\
     & ($\mu \rm m^3$)       & (N/m) &    (pg)  & & (MHz) & & ($\mu \rm $N) & ($ \rm m^2 \times 10^{-21}$) & ($ \rm m^2 \times 10^{-21}$) \\
\hline
$60 \mu \rm m$ &  $60 \times 0.95 \times 0.093$ &  0.93  & 10.82 & 0.476 & 1.48 & 720 & 7.64 & 4.36 & 4.41  \\
$50 \mu \rm m$ &  $50 \times 0.95 \times 0.093$ &  1.24 & 9.63 & 0.469 & 1.80 & 500 & 7.64 & 3.29 & 3.33   \\
$40 \mu \rm m$ &   $40 \times 0.95\times 0.093$ &  1.42 & 6.53 & 0.462 & 2.35 & 320 & 7.64 & 2.86 & 2.75   \\
$30 \mu \rm m$ &   $30 \times 0.95\times 0.093$ &  1.66 & 3.33 & 0.451 & 3.55 & 180 & 7.64 & 2.46 & 2.56 \\
$20 \mu \rm m$ &   $20 \times 0.95\times 0.093$ &  2.09 & 1.83 & 0.434 & 5.38 & 80 & 7.64 & 1.95 & 1.94  \\
$15 \mu \rm m$ &   $15 \times 0.95\times 0.093$ &  3.42 & 1.26 & 0.423 & 8.28 & 45 & 7.64 & 1.19 & 1.18  \\
\end{tabular}
\end{ruledtabular}
\end{table}

\subsubsection{Theoretical Description of the Natural Frequencies and Mode Shapes for a Beam with Tension}
\label{Beam Equation and Eigenmodes}
Due to the pre-stressed nature of the silicon nitride layer, the suspended beams are under tension. The tension causes the dynamics of the beams to deviate from that of a typical Euler-Bernoulli (EB) beam without tension and approach that of a string under tension. The dynamics can be described by adding a tension term to the EB beam equation, which can be used to find the eigenfrequencies and mode shapes~\cite{Bokaian1990,stachiv:2014} in the usual manner.

We will follow the analytical approach provided by Bokaian~\cite{Bokaian1990} to describe the dynamics of a doubly clamped-beam under tension in the absence of a fluid. The result of this analysis will be the  eigenfrequencies (natural frequencies) $\omega_n$ and mode shapes $\phi_n$ of the beam.  While we only need the mechanical parameters of the fundamental mode for the fluid experiments, we have used the higher mode frequencies to determine the material properties of  silicon nitride, as we describe in Section \ref{section:Determining Material Properties from Natural Frequencies}.

Our intention in the following is to provide only the necessary details needed to arrive at the expressions for   $\omega_n$ and $\phi_n$, and to clearly establish our notation and conventions. We refer the reader to Ref.~\cite{Bokaian1990} for more details. For a long and slender doubly-clamped beam that is under tension and undergoing small flexural oscillations the dynamics can be described by 
\begin{equation}
{\frac{E I}{l^4} \frac{\partial^4 {\cal W}({x^*},t)}{\partial {x^*}^4}} - {\frac{S}{l^2} \frac{\partial^2 { \cal W}({x^*},t)}{\partial {x^*}^2} }+ {\mu \frac{\partial^2 {\cal W}({x^*},t)}{\partial t^2}}  = 0.
\label{eq:beam-equation}
\end{equation}
The beam geometry is determined by its length $l$, width $b$, and thickness $h$. A long and slender beam indicates $l \gg b, h$, and small deflections indicate that the maximum beam deflection is much smaller than the thickness of the beam. Both of these conditions are satisfied by the nanomechanical devices that we explore here. In our notation, $x^* = x/l$ is the dimensionless axial coordinate along the beam where $x$ is the dimensional coordinate. In addition, ${\cal W}(x^*,t)$ is the transverse displacement (in the $z$ direction) of the beam at axial position $x^*$ and time $t$; $E$ is the Young's modulus, $I = b h^3/12$ is the area moment of inertia, $\mu = \rho_s b h$ is the mass per unit length of the beam, and $S$ is the applied axial force. The minus sign on the second term indicates that the beam is under tension. The boundary conditions for the doubly clamped beam are ${\cal W}(0,t)={\cal W}(1,t)=\frac{\partial {\cal W}(0,t)}{\partial x^*}=\frac{\partial {\cal W}(1,t)}{\partial x^*}=0$.

In the main text, we focus upon experimental measurements at a specific axial location on the beam ${x_0}^*$. To simplify the notation for this case, we define the time varying oscillations at location ${x_0}^*$ as $W(t) = {{\cal W}}({x_0}^*,t)$, where we have used ${x_0}^* = 1/2$. However, in the following we will keep the discussion general and use ${\cal W}(x^*,t)$.

We next assume that the oscillating solutions are of the form
\begin{equation}
{\cal W}({x^*},t) = {Y_n({x^*})} {e^{i \omega_n t}}
\end{equation}
where $n$ is the mode number, $Y_n({x^*})$ is the $n$\textsuperscript{th} mode shape, and $\omega_n$ is the $n$\textsuperscript{th} natural frequency. Substituting this expression into Eq.~(\ref{eq:beam-equation}) yields
\begin{equation}
{\frac{E I}{l^4} \frac{d^4 Y_n({x^*})}{d {x^*}^4}} - {\frac{S}{l^2} \frac{d^2 Y_n({x^*})}{d {x^*}^2}} - {\mu \omega_n^2 Y_n({x^*})} = 0,
\label{eq:beam-ode}
\end{equation}
whose solution can be expressed as
\begin{equation}
\label{eq:mode_shape}
{Y_n({x^*})} = {c_{1,n} \sinh(M_n {x^*})} + {c_{2,n} \cosh(M_n {x^*})} + {c_{3,n} \sin(N_n {x^*})} + {c_{4,n} \cos(N_n {x^*})},
\end{equation}
which is in terms of the constants $M_n = ( U + \sqrt{U^2 + {\Omega_n}^2})^{1/2}$ and $N_n = ( -U + \sqrt{U^2 + {\Omega_n}^2})^{1/2}$. The amount of tension in the beam is described by the nondimensional tension parameter $U$, where 
\begin{equation} \label{eq:U-defn}
    U= \frac{S}{2 E I/l^2} 
\end{equation} 
represents the ratio of the axial load on the beam to its rigidity. The nondimensional natural frequency for mode $n$ is $\Omega_n$, where 
\begin{equation}
    \Omega_n = \frac{\omega_n}{\beta/l^2} 
    \label{eq:Omega_n}
\end{equation}
and $\beta = (EI/\mu)^{1/2}$ is a constant determined by material properties and linear dimensions.

Boundary conditions, $Y(0) = Y(1) = d Y(0)/d{x^*} = d Y(1)/d{x^*} = 0$, allow for  determining the values of the constants as 
\begin{eqnarray}
c_{1,n} &=& {1}, \\ 
c_{2,n} &=&  {\frac{M_n \sin (N_n) - N_n \sinh (M_n)}{N_n \left[ \cosh(M_n) - \cos(N_n)\right]}}, \\
c_{3,n} &=& { - \frac{M_n}{N_n}},  \\ 
c_{4,n} &=& - c_{2,n}. 
\end{eqnarray}
The characteristic equation relating $U$ and $\Omega_n$ is 
\begin{multline}
{\Omega_n} + {U \sinh\left( U + \sqrt{U^2 + {\Omega_n}^2}\right)^{1/2}} {\sin\left( -U + \sqrt{U^2 + {\Omega_n}^2}\right)^{1/2}} \\ - {\Omega_n \cosh \left( U + \sqrt{U^2 + {\Omega_n}^2}\right)^{1/2}} {\cos \left( -U + \sqrt{U^2 + {\Omega_n}^2}\right)^{1/2}}  = {0}.
\label{eq:characteristic-equation}
\end{multline}
Given a value of $U$, all of the natural frequencies $\Omega_n$ are found as solutions of Eq.~(\ref{eq:characteristic-equation}). In practice, we determine the values of $\Omega_n$ numerically as the successive roots of Eq.~(\ref{eq:characteristic-equation}). We will find it convenient to use the normalized mode shapes $\phi_n({x^*}) = Y_n(x^*)/{\tilde{N}_n}^{1/2}$ where $\tilde{N}_n$ is chosen such that
\begin{equation}
{\int_{0}^{1}} {\phi_n({x^*})} {\phi_m({x^*})} {d{x^*}} = {\delta_{nm}}
\end{equation}
with $\delta_{nm}=1$ for $n=m$ and $\delta_{nm}=0$ for $n \ne m$. Note that the orthonormal eigenfunctions $\phi_n$ are dimensionless. 

\begin{figure}
\begin{center}
\includegraphics[width=3.375in]{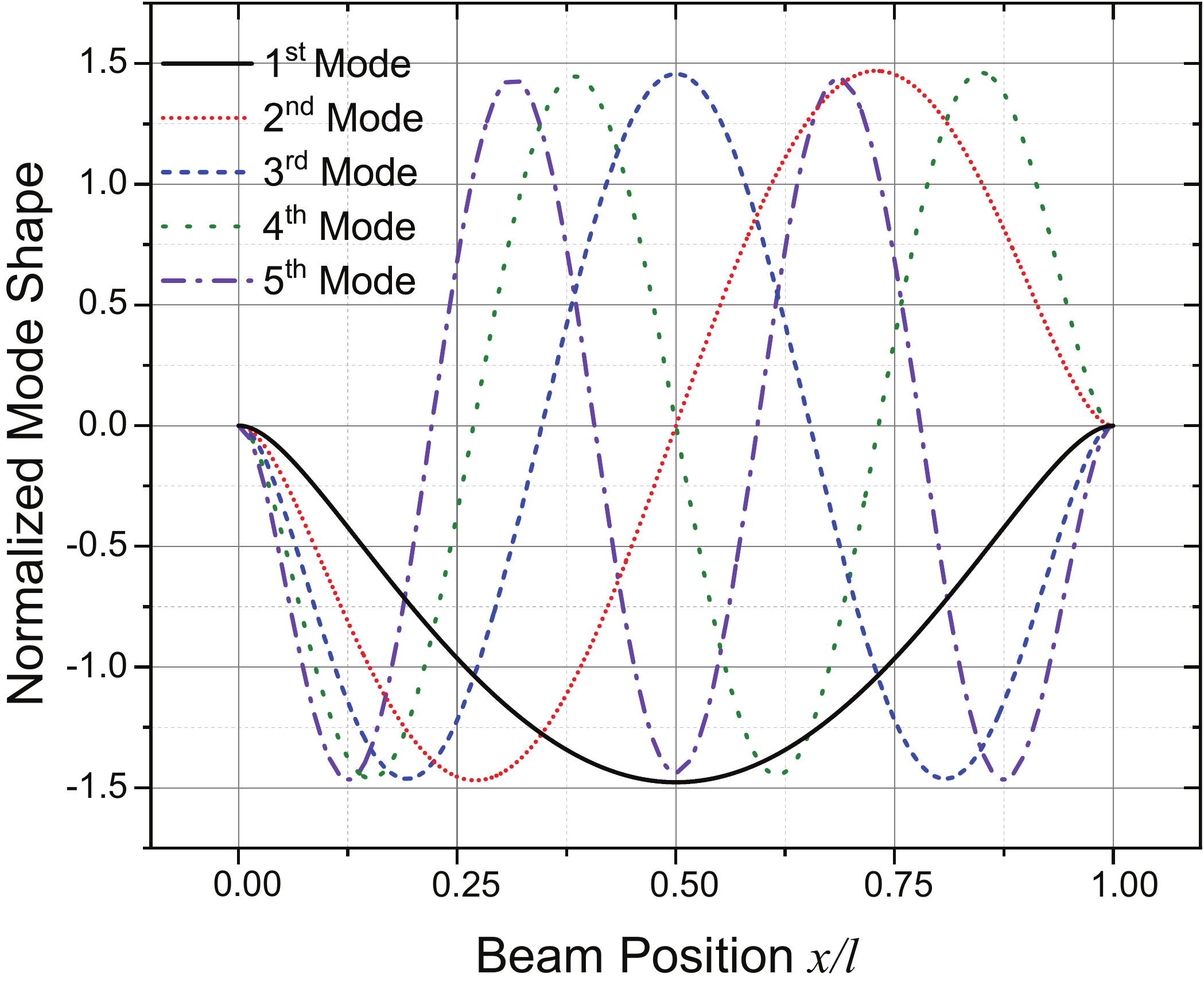}
\end{center}
\caption{The first five normalized mode shapes $\phi_n$ as a function of ${x}/{l}$ for a doubly-clamped beam under tension.  For this figure we have used $U=320$ which corresponds to the parameters of the 40-$\mu$m-long beam (Table~\ref{tab_devices} and \ref{fit_parameters}).}
\label{fig:modes}
\end{figure}

\subsubsection{Determining Material Properties from Natural Frequencies} \label{section:Determining Material Properties from Natural Frequencies}

Since we do not know the precise values of $S$, $E$ and $\rho_s$, we use an error minimization approach to determine these from theory and experimental observables of the beams in air, namely the eigenfrequencies. Theoretical predictions of the eigenfrequencies in terms of $\Omega_n$ and $U$ are given by Eq.~(\ref{eq:characteristic-equation}) where  $\Omega_n$ depends on  $\rho_s$ (Eq.~(\ref{eq:Omega_n})) and $U$ depends on $S$ and $E$ (Eq.~(\ref{eq:U-defn})). We take the following approach. We select trial values for $E$ and $\rho_s$. Using  the linear dimensions of the beam, the measured  $\omega_1$, and the trial values of $E$ and $\rho_s$, we compute the tension $S$ in the beam  by numerically solving Eq.~(\ref{eq:characteristic-equation}). We repeat this for all the fundamental frequencies of all the beams and find  average $S$ and $U$  values. Once $U$ is determined for these trial values of $E$ and $\rho_s$, we find all of the remaining eigenfrequencies  as the successive roots of Eq.~(\ref{eq:characteristic-equation}). Then, we form an error function defined as \begin{equation}
    \varepsilon (E, \rho_s) = \sum\limits_k \frac{|\omega_{k}^{(e)} - \omega_{k}^{(t)}|^2}{|\omega_{k}^{(e)}|^2},
\end{equation}
where $k$ runs through all  (available) modes of all the beams, and $\omega_{k}^{(e)}$ and $\omega_{k}^{(t)}$  are the experimental and theoretical eigenfrequencies. We repeat this process and calculate $\varepsilon$  by sweeping the  $\rho_s$-$E$ plane within the range of acceptable values from the literature which are $2600~ {\rm kg / m^3} \le \rho_s \le 3400 ~ {\rm kg / m^3}$ and $250~{\rm GPa} \le E \le 350 ~ {\rm GPa}$~\cite{Senturia:2002, Drummond:1995, Santucci:2001, Stoffel:1996, Larsen:2013, Fauver:1998, Petersen:1982, Edwards:2004, Yang:2008, Yang:2002, Tabata:1989}.  We find that $E= 300 ~\rm GPa$ and $\rho_s=2750~\rm kg/m^3$ values provide a global minimum for the function $\varepsilon (E, \rho_s)$ for the average tension value $S\approx 7.64 ~\rm \mu N$. This tension value corresponds to a stress of $\sigma \approx 86.5 ~\rm MPa$. These values thus provide useful approximations for the material proprieties of our silicon nitride sample. Table~\ref{table:frequencies} lists experimentally measured eigenfrequencies and the theoretical eigenfrequencies (given in parentheses).  The theoretical predictions of the frequencies are determined using the optimal values of $E$, $\rho_s$ and the average value of $S$.  

\begin{figure}
   \begin{center}
   \includegraphics[width=6.75in]{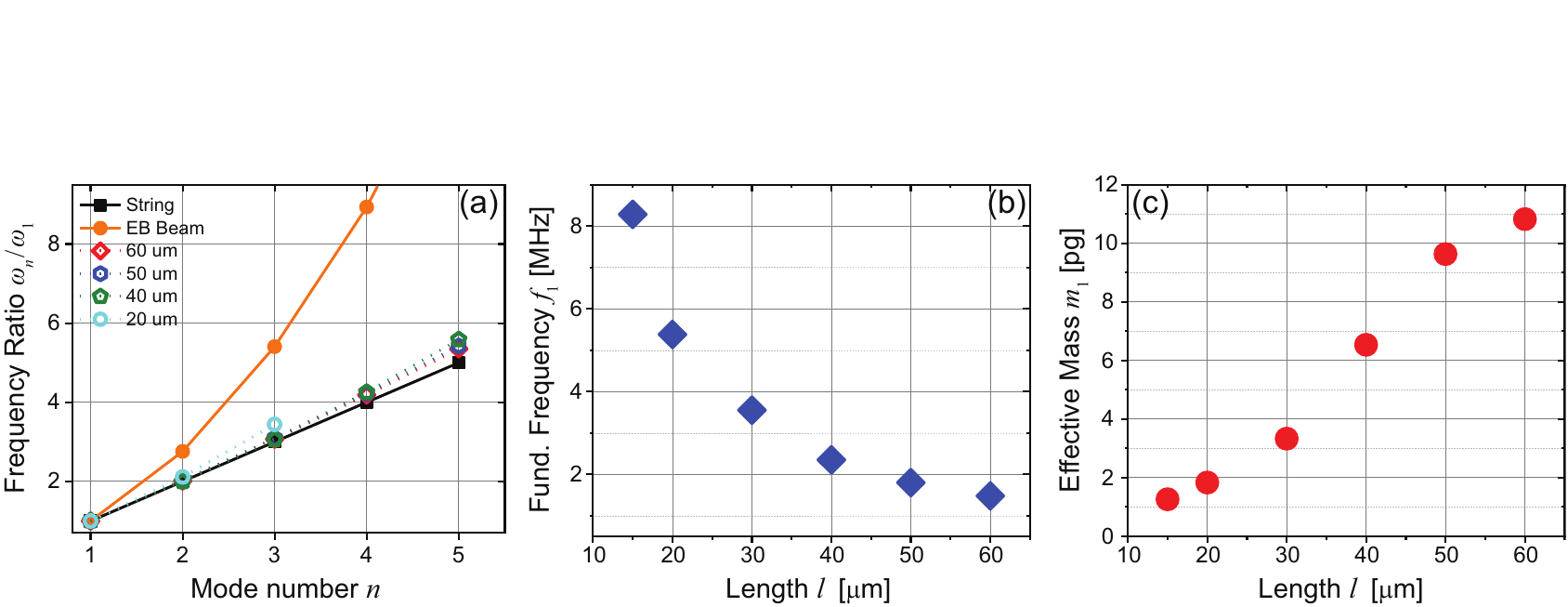} \\ 
    \caption{(a) Ratio of the $n^{\rm th}$ mode frequency to the fundamental mode frequency, $\omega_n/\omega_1$ for four of the devices used in the experiments. Theoretical ratios for an  Euler-Bernoulli beam without tension (orange) and a string under tension (black) are also shown. (b) Fundamental mode frequencies $f_1=\omega_1/2\pi$ and (c) effective masses $m_1$ for all the devices as a function of their lengths.}
    \label{fig:Supp_Tension}
    \end{center}
\end{figure}

The tension in the beams manifests itself in several ways.  First,  tension increases the natural frequencies of a beam resonator. Second, the relation between the resonance frequencies deviate from that of EB beam. Figure~\ref{fig:Supp_Tension}(a)  shows the ratio of $\omega_n/\omega_1$ for our beams as well as  the two  limiting cases. Orange symbols correspond to an EB beam where $\omega_n/ \omega_1\approx 1,2.75,5.40,\ldots$ for $n=1,2,3,\ldots$. The black symbols correspond to a string under tension where $\omega_n/\omega_1 = n$.  The ratios $\omega_n/\omega_1$ for all of our experimentally measured beams lie in between these two limits and are shown by the different open symbols.  This behavior is due to the fact that $U$ depends quadratically on the beam length as shown in Eq.~(\ref{eq:U-defn}).  

\begin{table}[b]
\caption{\label{tab_device_frequencies} Eigenfrequencies (natural frequencies) of the fundamental and higher modes of the devices  used in this study. Values in  parentheses are the theoretical estimations using the optimal material properties.}
\begin{ruledtabular}
\begin{tabular}{cccccccccc}
Device  & $\omega_1 / 2\pi$ &   $\omega_2 / 2\pi$ &  $\omega_3 / 2\pi$ &   $\omega_4 / 2\pi$ &  $\omega_5 / 2\pi$  \\
     &  (MHz) &  (MHz)  & (MHz) & (MHz) & (MHz) \\
\hline
$60 \mu \rm m$ &  1.48 (1.56)  & 2.93 (3.16) & 4.46 (4.82) & 6.10 (6.57) & 7.82 (8.44)  \\
$50 \mu \rm m$ &  1.80 (1.90) & 3.78 (3.86) & 5.78 (5.92) & 7.89 (8.15) & 10.14 (10.56)   \\
$40 \mu \rm m$ &  2.35 (2.42) & 4.66 (4.96) & 7.14 (7.70) & 9.94 (10.74) & 13.04 (14.13)   \\
$30 \mu \rm m$ &  3.55 (3.35) & - (6.96) & - (11.04) & - (15.77) & - (21.26) \\
$20 \mu \rm m$ &  5.38 (5.42) & 11.68 (11.68) & 18.67 (19.42) & - (28.94) & - (53.97)   \\
$15 \mu \rm m$ &   8.28 (7.85) & - (17.61) & - (30.39) & - (46.67) & - (66.64) \\
\end{tabular}
\end{ruledtabular}
\label{table:frequencies}
\end{table}

Finally, the modeshapes also depend on $S$, but  only weakly. The first five normalized mode shapes calculated for a 40-$\rm \mu m$-long beam are shown in Fig.~\ref{fig:modes}. We note that it is experimentally not possible for us to measure the modeshapes with enough precision such that we can  distinguish between an EB beam and a string under tension.  All of the relevant parameters used in finding the mode shapes are also given in Table~\ref{fit_parameters}.

\subsubsection{The Spring Constant of the Fundamental Eigenmode} \label{eigen}

\begin{figure}
   \begin{center}
   \includegraphics[width=4.5 in]{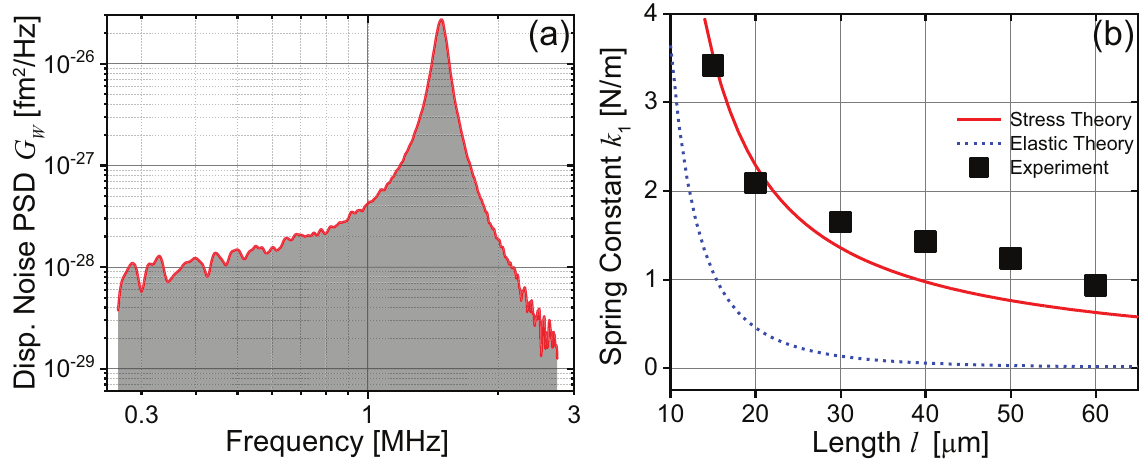}
    \caption{(a)~Thermal noise peak  of a 60-$\rm \mu m$-long nanomechanical beam resonator in air. The shaded area shows the integral of the PSD and is used in the equipartition theorem to calculate the spring constant of the device. (b) Experimentally measured spring constants $k_1$ as a function of the device length $l$ (symbols). The blue dotted curve is the theoretical prediction in the absence of tension given by Eq.~(\ref{elastic}). The red solid line  includes tension (Eq.~(\ref{lachut})).}
    \label{fig:Supp_Noise_integrate}
    \end{center}
\end{figure}

In the main text, we describe how the spring constant of the fundamental mode $k_1$ for each beam is determined using the classical equipartition theorem. Figure~\ref{fig:Supp_Noise_integrate}(a) shows an example of how we  calculate the mean-squared thermal displacement for the fundamental mode of a 60-$\rm \mu m$ device from the power spectal density  (PSD) of its Brownian noise in air. Since the thermal resonance is sharply peaked, we can find the area under the curve by integration, which provides the mean-squared displacement $\langle {W_{air}}^2 \rangle$. In Fig.~\ref{fig:Supp_Noise_integrate}(a), this area is the shaded region under the curve (red). Then, using the classical equipartition theorem, we find the spring constant as $k_1  = \frac { k_B T} {\langle {W_{air}}^2 \rangle}$, where $k_B$ is the Boltzmann constant and $T$ is the temperature. The variation of the measured spring constants $k_1$ with length $l$ for our devices are shown in Fig~\ref{fig:Supp_Noise_integrate}(b) by the data points. In addition, the  values for $k_1$ are listed in Table~\ref{tab_devices}.

As a comparison, the static spring constant of a doubly-clamped beam without tension when a point load is applied perpendicularly at the center of the beam, $x=l/2$, is given by~\cite{Cleland2003}
\begin{equation}
k_0 = \frac{192 E I}{l^3}.
\label{elastic}
\end{equation}
When the beam is under significant stress, Eq.~(\ref{elastic}) becomes less accurate, since the axial load must be taken into account. Lachut \textit{et al.}~\cite{Lachut2012} formulated a method to calculate the spring constant of a beam under stress, taking into account the axial load. For a beam with stress of $\sigma_s$,  the ratio of the increase $\delta k$ in the spring constant to the stress-free spring constant $k_0$ is given by 
\begin{equation}
\label{lachut}
{\frac{\delta k}{k_0} }= {\frac{3}{10}}{ \frac{\left(1 - \nu_{p}\right) \sigma_s}{E h}} {\bigg(\frac{l}{h}\bigg)^2},
\end{equation}
where $\nu_p$ is the Poisson ratio. The stress $\sigma_s$ is the result of the axial load $S$, where $S= (1 - \nu_p)\sigma_s b $. The spring constant is then given by $k=k_0 + \delta k$. Figure~\ref{fig:Supp_Noise_integrate}(b) includes the predicted   spring constants  using Eq.~(\ref{elastic}) (blue, dotted line) and using Eq.~(\ref{lachut}) (red, solid line), using the linear dimensions, material properties, and tension values in Table~\ref{fit_parameters} along with $\nu_p =0.28$.

\subsubsection{The Effective Mass of the Fundamental Eigenmode}
\label{subsection:effective_mass}

The effective mass $m_1$ of the fundamental mode of a beam can be found using $m_1=k_1/ {\omega_1}^2$. In modal analysis, the effective mass (and the spring constant) can be defined only up to a constant. However, since we experimentally measure the resonance frequencies $\omega_1$ and determine the spring constants $k_1$ using the equipartition theorem, there is no ambiguity in our values, and we are able to present quantitative measurements for $m_1$.  Figure~ \ref{fig:Supp_Tension}(c) shows the effective mass $m_1$ as a function of the beam length $l$. In addition, the numerical values of measured $m_1$ are  listed in Table~\ref{tab_devices}.

In our experiments we use a point measurement at the optical spot to quantify the beam dynamics. In the following, we describe how this point measurement is connected with our lumped single-degree-of-freedom model. We equate the kinetic energy of the fundamental mode as measured at position $x=x_0$ of the beam to the kinetic energy of the entire beam with physical mass $m=\rho_s l b h$ that is oscillating at frequency $\omega_1$ as~(cf.~\cite{Cleland2003,Cross2006}). This yields 
\begin{equation}
\label{eff_mass}
\frac{1}{2} m_1 \dot{{\cal W}}(x_0,t)^2 = \frac{1}{2} {\int_{0}^{l}} \mu \dot {\cal W}(x,t)^2 dx,
\end{equation}
where $m_1$ is the effective mass of the fundamental mode when measured at position $x_0$. If we assume $\mu = m/l$ is constant and describe the beam motion using the fundamental mode shape as ${\cal W}(x,t) = a_1 Y_1(x) e^{i \omega_1 t}$ where $a_1$ is an arbitrary constant determining the amplitude of the motion, this expression can be simplified to
\begin{equation}
    \frac{m_1}{m} = \frac{1}{ Y_1(x_0)^2 l} {\int_{0}^{l}} Y_1(x)^2 dx,
\end{equation}
which is equivalent to 
\begin{equation}
    \frac{m_1}{m} = \frac{1}{ Y_1({x_0}^*)^2} {\int_{0}^{1}} Y_1(x^*)^2 dx^*,
\end{equation}
where we now use the nondimensional coordinate $x^*$ and ${x_0}^* = x_0/l$. Using the normalized dimensionless mode shape $\phi_1$, this can be written as
\begin{equation}
    \frac{m_1}{m} = \frac{1}{\phi_1({x_0}^*)^2} {\int_{0}^{1}} \phi_1(x^*)^2 dx^*.
\end{equation}
Lastly, the integral is unity because of the orthogonality relationship for $\phi_n$, which gives the final result
\begin{equation}
    \frac{m_1}{m} = \frac{1}{ \phi_1({x_0}^*)^2}.
    \label{eq:moverm}
\end{equation}
In the following, we will define $\alpha_1 = 1/\phi_1({x_0}^*)^2$ which yields the following useful relationship between the effective mass of the fundamental mode when measured at location ${x_0}^*$ and the actual mass of the beam: $m_1 = \alpha_1 m$.

Since all of our experimental measurements are at the center of the beams, we will always consider $\alpha_1$ to be determined at ${x_0}^*=1/2$.  For an Euler-Bernoulli doubly-clamped beam without tension, $\alpha_1 = 0.3965$, and for a string under tension, $\alpha_1 = 1/2$. For the devices used in this work,  $0.423\le \alpha_1 \le 0.476$, corresponding to the shortest and longest beams, respectively. The numerical values for  $\alpha_1$ for all our beams are  provided in Table~\ref{fit_parameters}. These $\alpha_1$ values are calculated using the fundamental mode shapes of the beam with tension (Section \ref{Beam Equation and Eigenmodes}) using the tension value $S= 7.64~\rm \mu N$. 

We can theoretically predict a value of ${m_1}$ using ${m_1} = \alpha_1 m$, which requires the mode shape, beam geometry, and beam material properties. This theoretical value is also included in Table~\ref{fit_parameters} as the calculated effective mass. The difference between the experimentally measured mass and the calculated effective mass increases with beam length and is due to our limited knowledge of the precise device geometry and material properties. For example, the calculated effective mass is an estimation based on device dimensions and does not account for fabrication imperfections nor additional factors such as the masses of the metal films and the contributions of the overhangs at the base of the anchors (see Figure~\ref{fig:Supp_SEM}(b)-(c)).

\section{Measurements}

\subsection{Experimental Setup and Protocols}

\subsubsection{Optical Interferometer}

\begin{figure}
   \begin{center}
   \includegraphics[width=4.5in]{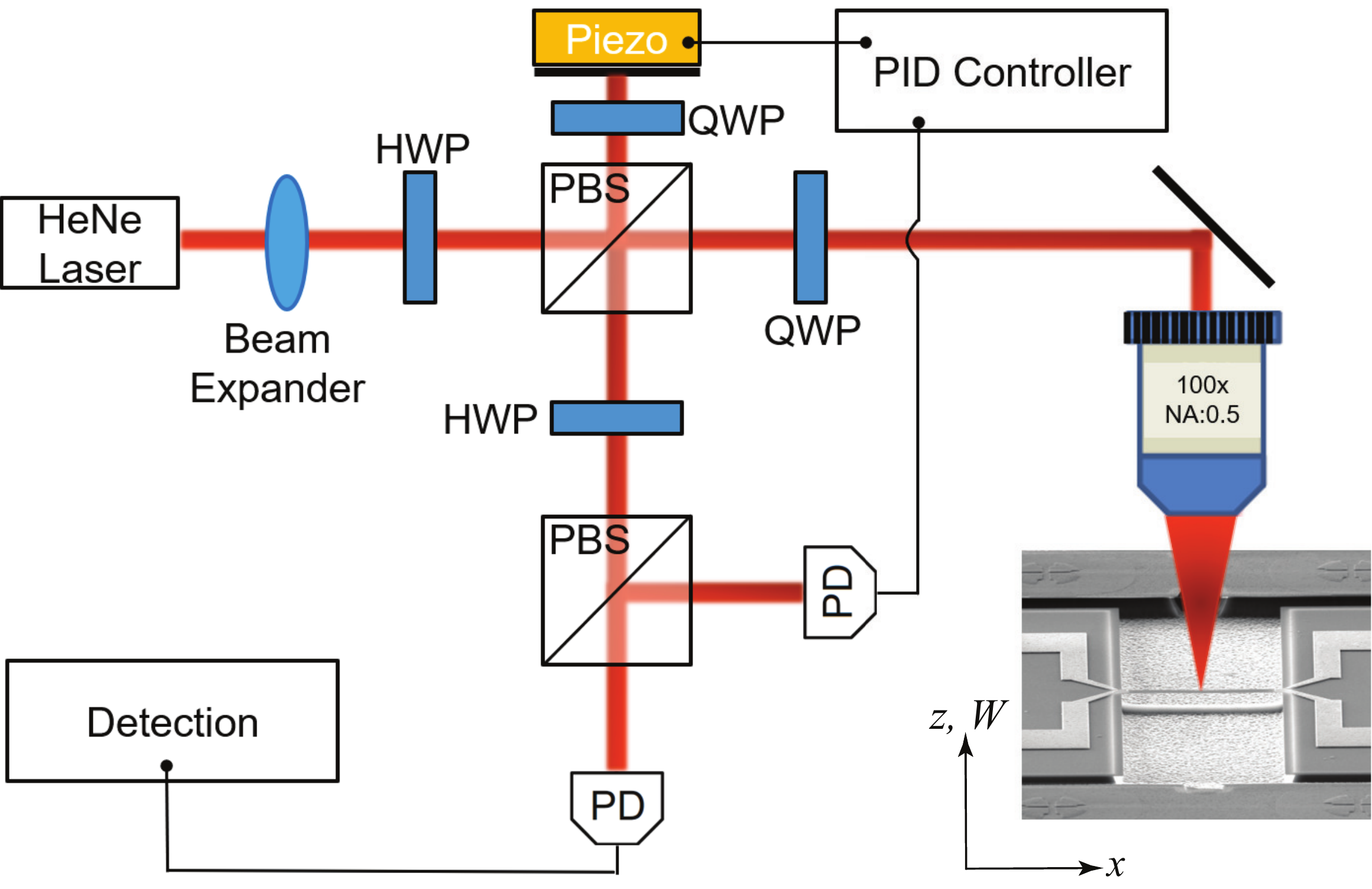} 
    \caption{Optical detection setup used for the experiments. HWP: half wave plate, QWP: quarter wave plate, PBS: polarized beam splitter, PD: Photodetector, PID: Proportional-integral-derivative controller. For thermal fluctuation measurements, the detection PD is connected to a spectrum analyzer; for driven measurements, it is connected to a lock-in amplifier. The second PD with a PID controller and a movable mirror is used for path stabilization.}
    \label{fig:Supp_Setup}
    \end{center}
\end{figure}

A homodyne optical interferometer with path stabilization is used for detecting the nanomechanical motion of the beams both in noise and driven measurements. A diagram of the optical setup can be found in Figure~\ref{fig:Supp_Setup}. A HeNe laser with a wavelength of $\lambda = 632.8~\rm nm$ and peak power of 1 mW is operated in its intensity stabilization mode. One photodetector (Thorlabs PDA8A) is used for path stabilization while a second photodetector (New Focus 1801) is used  to detect the oscillations of the beams. The displacement sensitivity of the interferometer is estimated to be $\sim25~\rm fm /\sqrt Hz$ in the range 1-50 MHz with {40 $\mu \rm W$} incident on the photodetector. In the noise measurements, we can further improve the displacement resolution to $\sim5~\rm fm /\sqrt Hz$ by averaging and numerically subtracting the background noise from the nanomechanical noise --- as discussed below in Section \ref{Noise_measurements}. The optical spot is $\sim800$ nm in diameter, and the typical power incident on a beam during a measurement is $~20 ~\mu \rm W$. 

\subsubsection{Experimental Protocols}

The devices are glued to a chip carrier; the measurements are performed with the chip carrier placed on a piezoelectric $XYZ$ stage. The laser, shown in Fig.~\ref{fig:Supp_Setup}, is focused on a  beam at the measurement location and the $XYZ$ stage is moved along the $x$ axis for the measurements. The  beam length is much larger than the optical spot radius, and the measurements are essentially at a single point on the beam. For liquid experiments, we fill the  chip carrier with the liquid (deionized water or isopropyl alcohol)  and then seal the  carrier  with a thin glass lid to prevent evaporation. We note that the depth of the liquid bath is much larger than any device dimensions. The liquid inside the cavity does not show any visible signs of evaporation even after several days.

\subsection{Noise Measurements} \label{Noise_measurements}

\begin{figure}[b!]
   \begin{center}
   \includegraphics[width=6.75in]{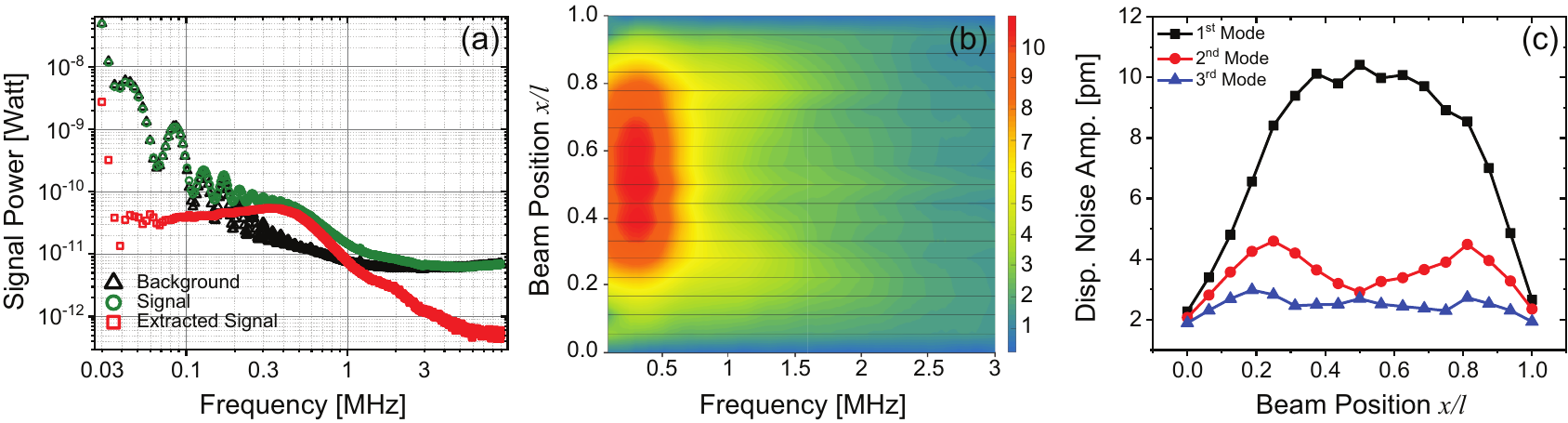} 
    \caption{(a) Background subtraction on noise data  of a 50-$\rm \mu m$ beam in water. The  background  (black) is subtracted from the measured nanomechanical noise (green) to find the subtracted noise (red). This subtracted noise is then converted to displacement noise by calibrating against the wavelength of the laser. Measurement bandwidth here is 20 $\rm kHz$.  (b) A color-map of the noise response of the of 50-$\rm \mu m$ beam immersed in water, showing the noise amplitude as a function of frequency and location of the measurement on the beam. In order to construct the color-map, we have collected noise measurements, such as those in (a), along the beam length at equal steps, starting from one anchor to the other. The data  are interpolated using Matlab's surf function. The color-bar corresponds to the root-mean squared displacement amplitude (within the 20 kHz bandwidth) in units of $\rm pm$. (c) Experimentally measured mode shapes of the first three out-of plane modes for the 50-$\rm \mu m$ beam. These are cross-sections from the color plot along the beam position at frequencies of 0.23 MHz, 0.63 MHz, and 1.41 MHz.}
    \label{fig:Supp_Noise}
    \end{center}
\end{figure}
In  noise measurements, a spectrum analyzer is used to detect the optical signals on the photodetector. We numerically subtract the background signal from the measurement signal to remove the low-frequency laser noise. Figure~\ref{fig:Supp_Noise}(a) shows an example of this background subtraction process. Brownian motion signals are recorded by focusing the laser on a location of interest on the nanomechanical beam (green trace in Fig.~\ref{fig:Supp_Noise}(a)). We average  $\sim10^4$ data traces for each measurement at a $20$-kHz resolution bandwidth. The measurements typically take several hours each. We then repeat the procedure with identical experimental parameters on the anchor of the beam on top of the silicon nitride layer to obtain the background signal (black trace in Fig.~\ref{fig:Supp_Noise}(a)). The assumption is that noise is uncorrelated and therefore noise powers can be added (or subtracted). The subtracted signal is then calibrated using the laser wavelength to find the PSD in units of $\rm m^2/Hz$. 

The frequency and position dependent  thermal  response of a 50-$\rm \mu m$-long nanomechanical beam is shown in Fig.~\ref{fig:Supp_Noise}(b). For this color-map in water, the PSD of the thermal fluctuations of the beam is measured along the beam length at equal intervals by moving the $XYZ$ stage.  The root-mean square (rms) amplitude at each point is found by multiplying the PSD with the resolution bandwidth and then taking the square root. The color-map is then generated using MATLAB's surf function. Figure~\ref{fig:Supp_Noise}(c) shows the measured mode shapes of the first three thermal modes of the  beam  in water along the beam length. These are essentially slices along the frequency axis in the color-map of Fig.~\ref{fig:Supp_Noise}(b) taken at frequencies of 0.23 MHz, 0.63 MHz, and 1.41 MHz. While noisy, the measured modeshapes  match the eigenmodes of the undamped beam shown in Fig.~\ref{fig:modes} closely.   We note that the  displacement noise measured in the experiment is reported as rms  and always remains  positive, in contrast to the theoretical mode shape. 

\subsection{Electrical Properties of the Electrothermal Actuators}

Electrothermal actuators are used to excite  the harmonic response  and measure the susceptibility of the nanomchanical beam resonators.  Bargatin \textit{et al.} \cite{Bargatin2007a} showed that the thermal time constant for such an actuator is on the order of ${\tau_{th}}^{-1} = {2\pi\times40~ \rm MHz}$. If one drives the actuators at shorter time scales  than ${\tau_{th}}$, the  actuation does not work efficiently. In our experiments the maximum drive frequency is 5 $\rm MHz$, well below this thermal cut-off frequency. Moreover, the above-mentioned time constant was determined in vacuum where the heat transfer from the actuator is dominated by conduction through the solid. In our experiments, the devices are  immersed in a liquid, where the medium enhances the  heat transfer and facilitates more rapid cooling. 
 
\begin{figure}
   \begin{center}
   \includegraphics[width=4.5in]{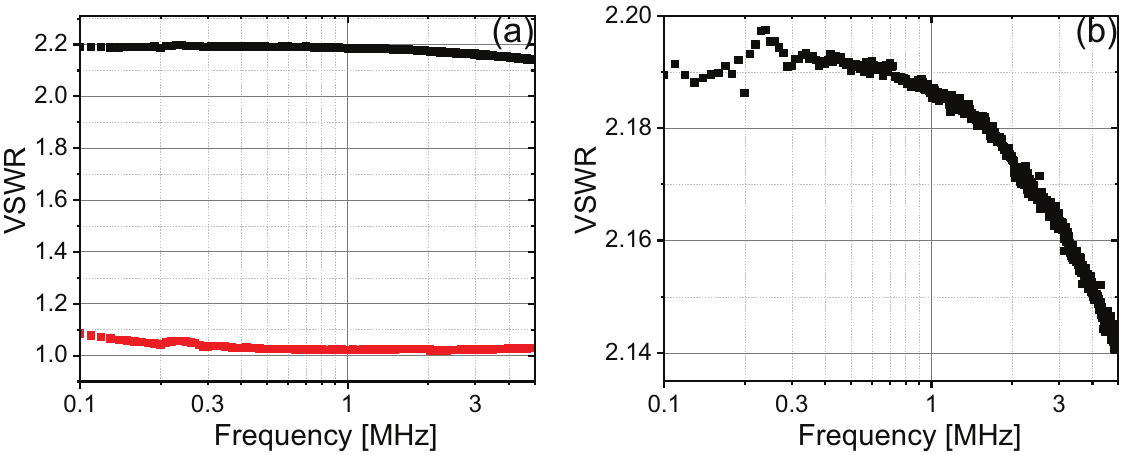} 
    \caption{(a) VSWR measurements for a 24.4 $\Omega$ electrode and a 50 $\Omega$ reference resistance in the range of $0.1-5~ \rm MHz$ plotted in a semi-logarithmic graph. (b) Close up of the transducer VSWR. The VSWR decreases from 2.19 to 2.14 in the measurement range which corresponds to  a $\le 1\%$ change in the dissipated power.}
    \label{fig:Supp_VSWR}
    \end{center}
\end{figure}

The electrothermal actuators are designed so that their resistances are close to $50~\Omega$. This ensures that the actuation power is dissipated in the resistor,  making the actuation process less susceptible to parasitic circuit elements. In order to confirm that the actuation power (and hence the force) remains constant over the frequency range of our experiments, we have measured the voltage standing wave ratio (VSWR) of the electrothermal actuators. The VSWR is typically expressed as \cite{Joines2012}
\begin{equation}
		{\rm VSWR} =\frac{1+\left | S_{11} \right|}{1-\left |S_{11}  \right |},
\end{equation}
where $S_{11}$ is the reflection coefficient and  quantifies the power reflected from the load. The  actuator electrical resistances throughout the devices are around $20\pm 5 ~\Omega$, with the uncertainty coming from  fabrication imperfections and design parameters. Figure~\ref{fig:Supp_VSWR}(a) shows the measured VSWR for an electrothermal transducer that is $24.4~\Omega$ and a $50~\Omega$ resistor for comparison, in the  frequency range ($0.1-5~ \rm MHz$) of   all the experiments. The VSWR drops from $\approx 2.19$ to $\approx 2.14$ over this frequency range, which corresponds to less than a 1\% change in the reflected power. Thus, the power dissipated on the transducer is, for all practical purposes, constant. 

The two-dimensional sheet resistance of the gold transducer layer is estimated to be $ R_s = 0.24~ \rm \Omega/ \square$, and the  resistance of the u-shaped actuator wire that sits close to the anchor of the beam is calculated as $\approx 3.5~\Omega$. A close-up top view of the actuator is shown in Fig.~\ref{fig:Supp_SEM}(c). The parasitic capacitance of the actuators have been estimated from other measurements of similar chips to be around $\sim 100 ~\rm pF$.

\subsection{Driven Measurements}

In the driven measurements, a signal generator drives the transducers with a sinusoidal voltage waveform. The frequency of the sinusoidal signal is swept while the amplitude is kept constant. To detect the signal, a lock-in amplifier is used in $2f$ mode. The power dissipated on a resistor due to an applied oscillating  voltage, $V_0 \cos \left({\frac {\omega_0t}{2}}\right)$, is proportional to $\frac{V_0^2}{2} +  \frac{ V_0^2}{2} \cos \omega_0 t $, where the constants $V_0$ and $\omega_0$ are the drive voltage and angular frequency, respectively.  The first term in the power dissipation expression causes an increase in the average temperature of the resistor and the second term causes the temperature to oscillate at twice the drive frequency. The temperature oscillations result in nanomechanical oscillations, as described in \cite{Bargatin2007a}. 

\begin{figure}
   \begin{center}
   \includegraphics[width=6.75in]{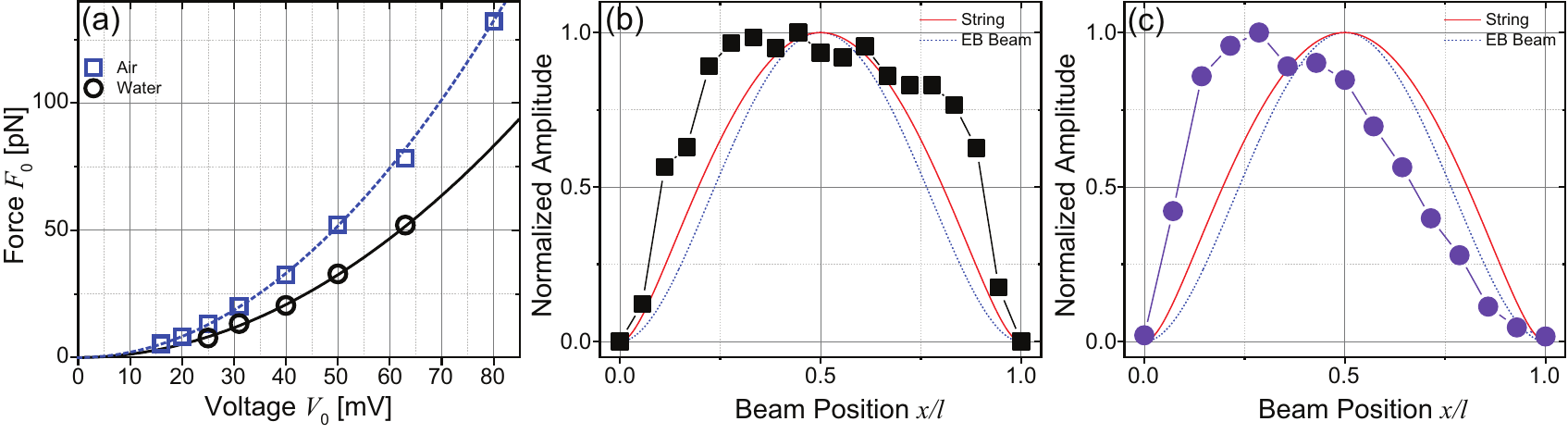} 
    \caption{(a)~Calculated drive force amplitude $F_0$ \textit{vs.} the drive voltage amplitude $V_0$ for the 60-$\mu \rm m$ beam in air (blue, squares) and in water (black, circles). Air calibration is based on the amplitude at the fundamental resonance frequency, whereas water calibration is based on the amplitude at $\omega \rightarrow 0$. The dashed blue and solid black lines are fits to $F_0 = A{V_0}^2$ with $A=2.06 \times 10^{-8}~\rm N/V^2$ and $A=1.30 \times 10^{-8}~\rm N/V^2$ for air and water, respectively. (b) The experimentally measured mode shape of the fundamental mode of a 50-$\mu \rm m$ beam in water (black, squares) that has been actuated at both ends of the beam. (c)~The experimentally measured mode shape of the fundamental mode of the 20-$\mu \rm m$ beam (violet, circles) driven by only one actuator near $x=0$. In panels (b)-(c), the red solid line is the theoretical mode shape of a string with tension and the blue dotted line is the theoretical mode shape of an Euler-Bernoulli beam with tension.}
    \label{fig:Supp_drive}
    \end{center}
\end{figure}

In the main text, we have calibrated the transducer in water using the response of the resonator at low frequency. This data set in water is shown in Fig.~\ref{fig:Supp_drive}(a)  as well as another data set that is taken  on the same resonator in air. In air, the drive data are calibrated using the amplitude at resonance --- taking advantage of the sharp resonance peak (see Fig.~1(b) in the main text). The amplitude of the mode at the resonance frequency $\omega=\omega_1$ can be expressed as ${{W}_{\omega_1} } = F_0 Q_1/k_1$. Here, the quality factor $Q_1$ of the peak has been found to be $Q_1 \approx 15$ from fitting the resonance lineshape to a Lorentzian.  With $k_1$, $Q_1$ and ${{W}_{\omega_1} }$ available from experiments, we determine the lumped force $F_0$ acting on this mode for each drive voltage $V_0$. The results are shown in Fig.~\ref{fig:Supp_drive}(a).

Both force calibration data sets in Fig.~\ref{fig:Supp_drive}(a) depend on the square of the applied voltage, as expected. The efficiency of the transducer is slightly higher in air, i.e., the anchors heat up to a higher temperature in air compared to water. This is  reasonable because the   thermal conductivity is higher in water. However, given that that there is no dramatic increase in transducer efficiency going from water to air, we conclude that  the   thermal conduction in these structures mostly  takes place through the solid.

Even though the structures are heating up to approximately the same temperature in air and water,  the oscillation amplitudes in water are significantly smaller than those in air  because of the viscous damping. We also note that the damping  is larger for longer beams. In order to alleviate this, the longer beams (40, 50, and 60~$\mu m$) are driven using  both transducers on the two ends of the beam as shown in Fig.~\ref{fig:Supp_SEM}(b) \cite{Bello:2020}. Figure~\ref{fig:Supp_drive}(b) shows the fundamental mode shape of a  50-$\rm \mu m$-long beam measured when both actuators are driving the nanomechanical motion. Here, the mode shape appears symmetric and close to the theoretical predictions. When only one transducer is used, the mode shape becomes slightly distorted toward the end of the beam where the drive transducer is located. An example of a distorted fundamental mode shape  of on a 20-$\rm \mu m $ beam is shown in Fig.~\ref{fig:Supp_drive}(c). This distortion is due to fact that the waves generated by the single transducer decay along the length of the beam because of the high viscosity before they can form the familiar standing wave patterns.  While this mode shape distortion is not significant for shorter beams, we anticipate it to be  a source of error when comparing the theory and experimental results. We  discuss this and other sources of errors further in Section~\ref{section_three}.

\subsection{ Measurements in Isopropyl Alcohol (IPA)}

\begin{figure}
\begin{center}
\includegraphics[width=4.5in]{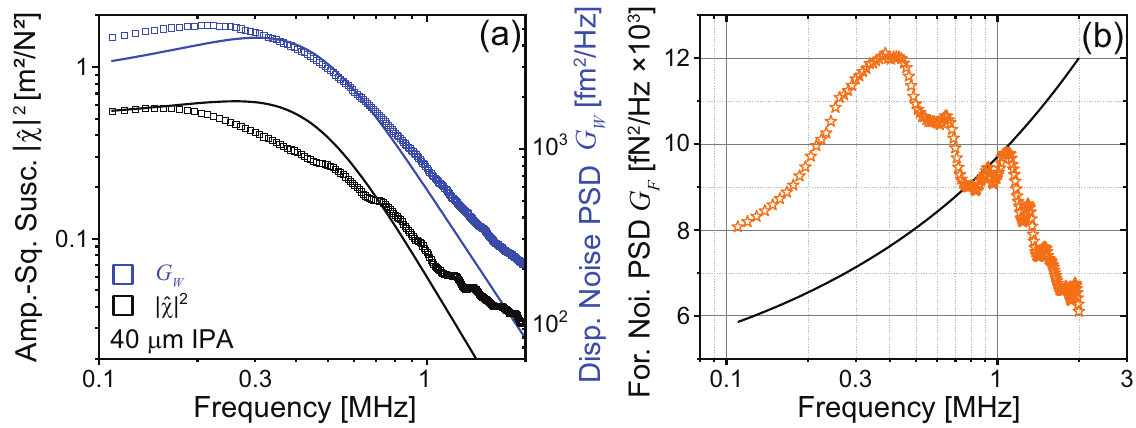}
\end{center}
\caption{(a) Amplitude-squared susceptibility $\left|{\hat \chi (\omega )} \right|^2$ (black) and PSD of the Brownian fluctuations (blue) for a 40-$\mu \rm m$ beam in IPA. Both quantities are measured at the center of the beam.  The symbols show the experimental data, and the continuous lines are predictions of the single-mode theory with no free parameters. (b) PSD of the Brownian force acting on the beam in IPA. The continuous lines show the normalized experimental thermal (blue) and driven (black) responses for each beam from (a) as well as the theoretical PSD of the Brownian force (black).}
\label{fig:IPA}
\end{figure}

In addition to  water  measurements, we have measured the Brownian force PSD for a 40-$\mu \rm m$ beam immersed in isopropyl  alcohol (IPA) as shown in  Fig.~\ref{fig:IPA}(a). IPA is more viscous than water ($\nu_f = 2.78 \times 10^{-6}~ \rm m^2/s $) but has a lower density ($\rho_f =786~ \rm kg/m^3$). Thus, the mechanical response of a given resonator is more damped in IPA  as compared to its response in water. This can be seen by comparing  Fig.~\ref{fig:IPA}(a) with Fig. 2(c) in the main text.  The experimental and theoretical curves for the PSD of the Brownian force $G_F$ are shown in Fig. \ref{fig:IPA}(b). The theoretical estimations are found using Eqs.~(\ref{G_d_supp}) and~(\ref{chi_d_supp}) which are discussed further below. All of the necessary fit parameters are included in Table~\ref{fit_parameters}.

\subsection{Mass Loading and Frequency Shift}

\begin{figure}
\begin{center}
\includegraphics[width=3.375in]{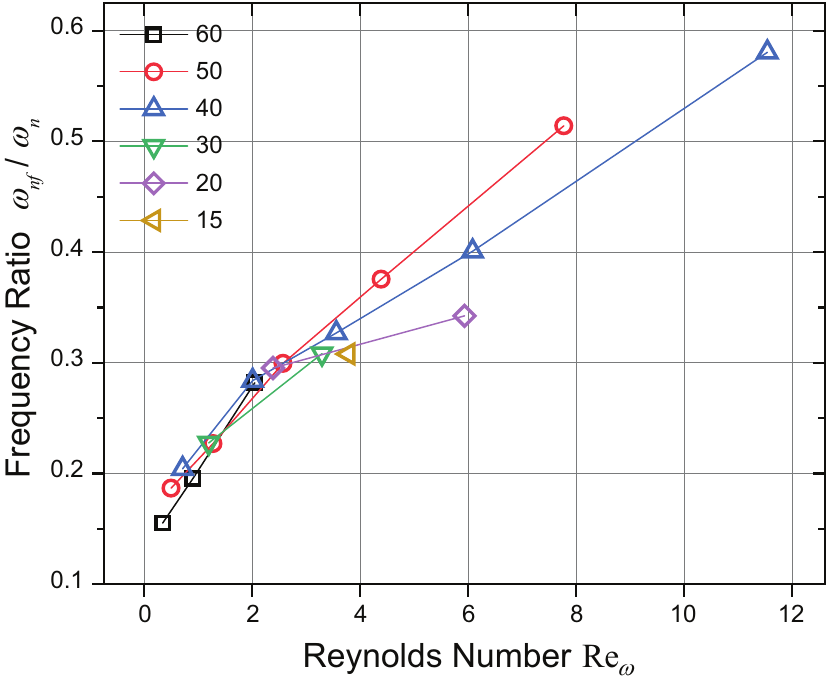}
\end{center}
\caption{ Ratio of the $n^{\rm th}$ mode frequency  in water ($\omega_{nf}$) to that in air ($\omega_{n}$) for each beam as a function of frequency-dependent Reynolds number, where ${\rm Re}_\omega= \frac{\omega_{nf}b^2}{4\nu_f}$.}
\label{fig:omega_water}
\end{figure}

In the fluid-structure interaction problem discussed below in Section \ref{subsection:cylinder_approximation}, one obtains an in-phase solution that couples to the mass term and gives mass loading.  This added mass  depends upon the  oscillation frequency and decreases the frequency of the peak in thermal and driven measurements in fluids. In Fig.~\ref{fig:omega_water}, we show our measurements of the peak frequencies of different beams and modes in water, all obtained from driven responses. The frequency of each peak in water,  $\omega_{nf}$, is normalized by the frequency of the same peak in air, $\omega_n$. The data are presented as a function of the frequency-dependent Reynolds number ${\rm Re}_\omega$, where  ${\rm Re}_\omega$  is  based on $\omega_{nf}$ (see Eq.~(\ref{Reynolds}) below). Because all these beams have the same width, ${\rm Re}_\omega$ is simply proportional to $\omega_{nf}$. In the region $0<{\rm Re}_\omega \le 4$,  the data plot allows us to estimate any unknown $\omega_{nf}$ accurately.

\section{Theory}
\label{section_three}

In this section, we briefly derive the expressions that describe the dynamics of the beam with tension in a viscous fluid. We provide our results in terms of a  lumped single-degree-of-freedom description. The theoretical description discussed in detail in Refs.~\cite{Cross2006,Clark2010} can be directly applied to our case with only a few modifications. In essence, the type of beam and the presence of tension affect the mode shape and the frequency of oscillation. Once the mode shape is known, it is possible to determine the fundamental natural frequency $\omega_1$ and the parameter $\alpha_1$ which can be used to determine the effective mass $m_1$ and spring constant $k_1$. With $\omega_1$, $m_1$ and $k_1$ determined, the previously-derived expressions~\cite{Cross2006,Clark2010} can be used to obtain the PSD of the displacement fluctuations of the beam and the susceptibility. 

\subsection{A Theoretical Description of the Fundamental Mode}
\label{1DOF}

A point measurement made at a specific location on the beam can be described as a lumped single-degree-of-freedom system. In our work, we have  made  measurements in the middle ($x=l/2$) of the beam. In this description, the nanomechanical beam in its fundamental mode probed at its center can be described as a particle of effective mass $m_1$ with modal displacement (or amplitude) ${W}(t)$ that is attached to a linear spring of stiffness $k_1$. Furthermore, the beam is exposed to an externally-applied driving force $F_d(t)$ and to a force due to the surrounding viscous fluid $F_f(t)$. This can be expressed as
\begin{equation}
	m_1\ddot{W} + k_1 W = F_f(t) + F_d(t),
\label{eqm}
\end{equation}
where $k_1=m_1 {\omega_1}^2$. In order to compute the susceptibility $\hat \chi(\omega)$, we use an unit impulse for the drive force $F_d(t) = \delta(t)$ and transform into Fourier space to obtain 
\begin{equation}
- m_1 \omega^2 \hat{W}(\omega) + k_1 \hat{W}(\omega) = \hat{F}_f(\omega) + 1,
\end{equation}
where we have used the conventions
\begin{equation}
	{{W} (t)} = {{1} \over {2 \pi} }\int_{-\infty}^{\infty} \hat{W}(\omega)e^{-i \omega t} d\omega,
	\label{Fourier1}
	\end{equation}
\begin{equation}
	{\hat{W}(\omega)} = {\int_{-\infty}^{\infty}} {{W}(t)e^{i \omega t}} {dt}.
	\label{Fourier2}
\end{equation}
If the fluid force is separated into its real and imaginary components, these will  contribute to the mass and damping terms, respectively, which we can represent as 
\begin{equation}
	-{m_f(\omega)\omega^2\hat{W}(\omega)} - {i\omega\gamma_f(\omega)\hat{W}(\omega)} + {k_1\hat{W}(\omega)} = 1,
	\label{eqm_Fourier}
\end{equation}
where $m_f(\omega)$ is the frequency-dependent mass that includes the contribution from the fluid and $\gamma_f(\omega)$ is the frequency-dependent damping due to the fluid. Specific expressions for the oscillating cylinder and oscillating blade approximations are discussed in Section~\ref{subsection:cylinder_approximation}. Solving for $\hat{W}(\omega)$ and noting that this is the Fourier transform of the susceptibility, we have the desired result 
\begin{equation}
\hat{\chi}(\omega) = \frac{1}{k_1 - m_f(\omega )\omega ^2 - i\omega \gamma _f(\omega )}.
\label{x_w}
\end{equation}
The  squared magnitude of the susceptibility is then
\begin{equation}
|\hat{\chi}(\omega)|^2 = \frac{1}{\left[k_1 - m_f(\omega )\omega ^2\right]^2 + \left[\omega \gamma _f(\omega)\right]^2}.
\label{xw2}
\end{equation}
The PSD of the beam fluctuations when driven by Brownian motion is directly related to the susceptibility by~\cite{Cross2006}
\begin{equation}
G_{W}(\omega) = \frac{4 k_B T}{\omega} {{\hat \chi}''}(\omega),    
\end{equation}
where ${{\hat \chi}''}(\omega) $ is the imaginary component of $\hat{\chi}(\omega)$. This yields
\begin{equation}
G_{W} (\omega) = \frac{4 k_B T \gamma_f}{(-m_f \omega^2 + k)^2 + (\omega \gamma_f)^2}.
\label{eq:gw}
\end{equation}
By the the fluctuation-dissipation theorem, the PSD of the force noise (Brownian force) driving the beam can be expressed as 
\begin{equation}
G_F(\omega) = 4 k_B T \gamma_f(\omega),    
\end{equation}
which results in 
\begin{equation}\label{eq:SHO_noise_Gw_GF}
G_{W} (\omega) = \frac{G_F(\omega)}{(-m_f \omega^2 + k_1)^2 + (\omega \gamma_f)^2}.
\end{equation}
Lastly, we note that Eq.~(\ref{eq:SHO_noise_Gw_GF}) can also be expressed as
\begin{equation}
G_{W} (\omega) = |\hat{\chi}(\omega)|^2 G_F(\omega).
\end{equation}
Therefore, for the single mode description we have the result
\begin{equation}
G_F (\omega) = \frac{G_{W}(\omega)}{|\hat{\chi}(\omega)|^2}.
\end{equation}
In the main text, we use the ratio of the experimentally measured $G_W(\omega)$ and $|\hat{\chi}(\omega)|^2$ to explore the spectral properties of $G_F (\omega)$.

\subsection{Treating the Beam as an Oscillating Blade in Fluid}
\label{subsection:cylinder_approximation}

In order to find the frequency dependence of the damping and  mass from the last section, we look to fluid dynamics, i.e., we seek to evaluate the fluid force acting on the oscillating beam. We start by making the following assumptions: the fluid  is incompressible; the beam is long and thin ($l \gg b, h$); and its displacements are small compared to any of its linear dimensions. With these assumptions, the beam  can be approximated as an infinitely long cylinder oscillating transverse to its axis with the same displacement amplitude as the beam. The radius of the cylinder is taken as $b/2$ in order to seamlessly transition back to the beam solution. The mathematical treatment of the oscillating cylinder problem is provided in \cite{Rosenhead}, see also Refs.~\cite{sader:1998,Cross2006} for its application to oscillating beams. 

The hydrodynamic force acting on the cylinder is proportional to its oscillatory velocity, ${\omega} \hat{W}$, and can be written as~\cite{sader:1998,Cross2006}
\begin{equation}
   {{\hat F}_f}(\omega ) = m_{cyl,e}{\omega ^2}  \Gamma_c ({\rm{R}}{{\rm{e}}_\omega}) \hat{W}(\omega ). 
\label{eq:fluid_force}
\end{equation}
Here, ${\Gamma_c}({\rm{R}}{{\rm{e}}_\omega })$  is the complex hydrodynamic function for the oscillating cylinder and  has the form
\begin{equation}
    {\Gamma}_{c}({\rm{Re}_{\omega}}) = {1} + \frac{ {4 i K_1 (-i \sqrt{i \rm{Re_{\omega}}})}}{\sqrt{i \rm{Re_{\omega}}} K_0(- i \sqrt{i \rm{Re_{\omega}}})},
\label{Hydrodynamic}
\end{equation}
where $K_0$ and $K_1$ are zeroth and first order modified Bessel functions of the second kind. The argument of $\Gamma_c$ is expressed in terms of the frequency-dependent Reynolds number
\begin{equation}
     {{\rm Re}_{\omega}} = \frac{\omega b^2}{4 \nu_f},
\label{Reynolds}
\end{equation}
where $\nu_f$ is the kinematic viscosity of the fluid and $\rm Re_\omega$ plays the role of a nondimensional frequency in the theory.  The mass term $m_{cyl,e}$ is the effective mass of the oscillating cylinder of fluid of diameter $b/2$ where $m_{cyl,e} = \alpha_n m_{cyl}$ with $m_{cyl} = \rho_l \frac{\pi}{4} b^2 l$ the actual mass of the fluid cylinder. The coefficient $\alpha_n$ is described in Section~\ref{subsection:effective_mass}. Since we are focusing on mode 1, we set $n=1$ from here on.

To account for the rectangular cross-section of the beams, we further apply a correction factor to $\Gamma_c(\rm {Re_{\omega}})$ and find the hydrodynamic function of an oscillating blade $\Gamma_b$ immersed in a liquid~\cite{sader:1998}
\begin{equation}
     \Gamma_b(\rm {Re_{\omega}}) = {{\tilde{\Omega}} (\rm {Re_{\omega}}) {\Gamma_c}(\rm Re_{\omega})},
\label{correction}
\end{equation}
where $\tilde{\Omega}$ is a complex valued correction factor that is tabulated in Ref.~\cite{sader:1998}.
\begin{figure}
   \begin{center}
   \includegraphics[width=3.375in]{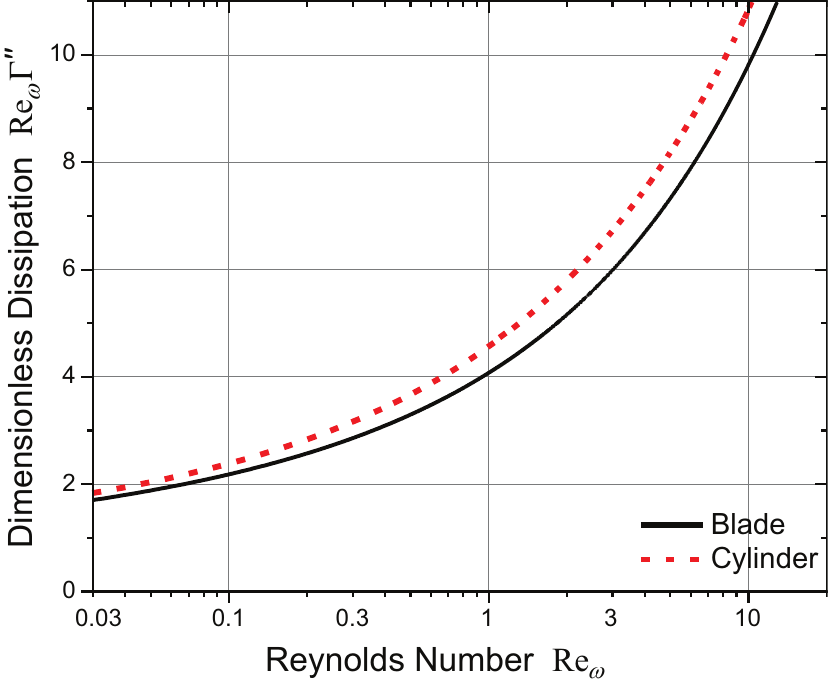} 
    \caption{Dimensionless PSD of the Brownian force as a function of the frequency-based Reynolds number ${\rm Re}_\omega$.  Eq.~(\ref{supp_GF_eq}) can be manipulated to show that $G_F$ $\propto {\rm Re}_\omega \Gamma''(\rm {Re_{\omega}})$. Theoretical predictions for the cylinder (red dashed line) and blade (black solid line) descriptions are shown.}
    \label{fig:Supp_Gf}
    \end{center}
\end{figure}

Separating ${\hat{F}_f(\omega)}$ to its real and imaginary parts, as we did in Eq.~(\ref{eqm_Fourier}), we express the frequency dependent mass and damping as~\cite{Cross2006}
\begin{equation} \label{eq:mass_1}
	m_f(\omega) = m_1 + m_{cyl,e} = m_1 + \alpha_1 {\rho_f \frac{\pi}{4} b^2 l {\Gamma}_b'(\rm {Re_{\omega}})} 
\end{equation}
and
\begin{equation} \label{eq:damping_1}
	{\gamma_f(\omega)} = \alpha_1 \rho_f {\frac{\pi}{4} b^2 l \omega {\Gamma}_b''(\rm {Re_{\omega}})}.
\end{equation}
The mass loading parameter $T_0$ is the ratio of the mass of a cylinder of fluid of diameter $b$ to the actual mass of the beam which can be expressed as
\begin{equation}  
	T_0 = \frac{m_{cyl}}{m_b} = {\frac{\pi}{4} \frac{\rho_f b}{\rho_s h}},
	\label{eq:mass-loading}
\end{equation}
where $\rho_f$ and $\rho_s$ are fluid and solid densities, respectively. Substituting into Eqs.~(\ref{eq:mass_1}) and (\ref{eq:damping_1}), we  write
\begin{equation}
	\label{m_f}
	 {m_f}(\omega ) = {m_1}\left( {1 + {T_0}{\Gamma _b'}({\rm{R}}{{\rm{e}}_\omega })} \right)
\end{equation}
and		
\begin{equation}
    \label{gamma_f}
	{\gamma_f(\omega)} = {m_1 T_0 \omega \Gamma_b''(\rm {Re_{\omega}})}.
\end{equation}
The PSD of the Brownian force $G_F$ then becomes
\begin{equation}\label{supp_GF_eq}
    G_F(\omega) = 4 k_B T  m_1 T_0 \omega \Gamma_b''({\rm Re}_\omega).
\end{equation}
The PSD of the Brownian force is proportional to ${\rm Re_{\omega}} \Gamma_b''(\rm {Re_{\omega}})$ for the cylinder and blade approximations and is shown in Fig.~\ref{fig:Supp_Gf}. Substituting the PSDs of the thermal force noise  $G_F$, frequency dependent mass $m_f$, and damping $\gamma_f$ into Eq.~(\ref{eq:gw}) and (\ref{xw2}), we arrive at explicit expressions for the PSD of Brownian fluctuations of the beam 
\begin{equation} \label{G_d_supp}
    {G_{W}(\omega)} = \frac{4 k_B T m_1 T_0 \omega \Gamma_b''({\rm Re}_\omega)}{{{{\left[ {{k_1} - {m_1}\left( {1 + {T_0}\Gamma_b'({\rm Re}_\omega)} \right){\omega ^2}} \right]}^2} + {{{\omega ^2} \left[ {{m_1}{T_0}\omega \Gamma_b''({\rm Re}_\omega)} \right]}^2}}}
\end{equation}
and for the squared-magnitude of the susceptibility 
\begin{equation} \label{chi_d_supp}
    {\left| {\hat\chi (\omega )} \right|^2} = \frac{1}{{{{\left[ {{k_1} - {m_1}\left( {1 + {T_0}\Gamma_b'({\rm Re}_\omega)} \right){\omega ^2}} \right]}^2} + {{{\omega ^2} \left[ {{m_1}{T_0}\omega \Gamma_b''({\rm Re}_\omega)} \right]}^2}}}.
\end{equation}
These expressions can be cast in a slightly different form as 
\begin{equation}
G_{W}(\omega) = \frac{1}{k_1^2} \frac{ 4 k_B T \alpha_1 m_b T_0 \omega \Gamma_b''(\omega)}{\left[1 - \tilde{\omega}^2 (1 + T_0 \Gamma_b'(\omega)) \right]^2 + \left[\tilde{\omega}^2 T_0 \Gamma_b''(\omega)\right]^2}.
\end{equation}
and
\begin{equation}
|\hat{\chi}(\omega)|^2 = \frac{1}{k_1^2} \frac{1}{\left[1 - \tilde{\omega}^2 (1 + T_0 \Gamma_b'(\omega)) \right]^2 + \left[\tilde{\omega}^2  T_0 \Gamma_b''(\omega)\right]^2}
\end{equation}
where $\tilde{\omega} = \omega /\omega_1$ to yield the expressions provided in~\cite{Cross2006}. 

In experiment, the beam is driven thermoelastically. The drive can be modeled as a sinusoidally varying force applied near the ends of the beam as 
\begin{equation}
F_d(x,t) = \begin{cases} \frac{F_0}{l} \sin (\omega_d t ) ~~~~ 0 \le x \le \xi_L ~~ \text{and} ~~  \xi_R \le x \le 1\\ 
                 0 ~~~~~~~~~~~~~~~~ \text{otherwise}
\end{cases}
\label{eq:driveforce}
\end{equation}
where $\xi_L$ and $\xi_R$ indicate the spatial extent over which the drive is applied. 
In our experiments we have used $\xi_L = 0.05$ and $\xi_R = 0.95$. Following the approach described in Ref.~\cite{Clark2010} the amplitude response can be expressed as
\begin{equation}
W_{\omega}(\omega)^2 =  \left( \frac{\pi F_0 \psi_1}{\phi_1(x_0)} \right)^2 \frac{1}{k_1^2} \frac{1}{\left( 1 - \tilde{\omega}^2 \left( 1 + T_0 \Gamma_b' \right) \right)^2 +  \left( \tilde{\omega}^2 T_0 \Gamma_b'' \right)^2 }
\end{equation}
where
\begin{equation}
\psi_1 = \int_0^{\xi_L} \phi_1(x/l) dx + \int_{\xi_R}^1 \phi_1(x/l) dx
\end{equation}
which captures the coupling of the driving force to the fundamental mode. For the single mode description we can express this result as
\begin{equation}
{W_{\omega}}^2 =  \left( \frac{\pi F_0 \psi_1}{\phi_1(x_0/l)} \right)^2 \left| \hat{\chi}(\omega) \right|^2,
\end{equation}
where it is apparent that the square of the amplitude is proportional to the magnitude of the susceptibility squared.

\subsection{Discussion and Possible Sources of Error}

The quantitative values used in our study are given in Tables~\ref{tab_devices},~\ref{tab_device_frequencies}, and~\ref{fit_parameters}. In Table~\ref{fit_parameters}, we have gathered together many of the important parameters used in our study for the purposes of further discussion. The top section of Table~\ref{fit_parameters} includes the material properties of the beam and the surrounding fluids that we have used.  To recapitulate, the Young's modulus $E$, the density of the silicon nitride beam  $\rho_s$, and the tension force in the beam $S$ are found using the optimization approach described in Section~\ref{section:Determining Material Properties from Natural Frequencies}. This provides a single set of values of $(E,\rho_s,S)$ that we use to describe all of the beams  in our study.  The simplification of using a single set of optimized values to describe all of the beams is expected to result in some error in the parameter values for an individual beam.

The middle section of Table~\ref{fit_parameters} includes several important nondimensional parameters. The mass loading parameter $T_0$ is evaluated using Eq.~(\ref{eq:mass-loading}). The mass loading parameter increases linearly with the density of the surrounding fluid. This is evident by the increase in $T_0$ when using water as the surrounding fluid when compared with the $T_0$ value for IPA. The nondimensional tension parameter $U$ is evaluated using Eq.~(\ref{eq:U-defn}). Since $S$ is a constant for our study, the value of $U$ increases quadratically with the length of the beam $l$. We have also included several values of the frequency based Reynolds number $Re_\omega$ given by Eq.~(\ref{Reynolds}). The Reynolds number provides insight into the relative importance of inertial and viscous effects. Each value of the Reynolds number is evaluated at the frequency of the peak of the fundamental mode $\omega_{1f}$ of the PSD of the Brownian fluctuations for that fluid. For air we use $\omega_1$ since $\omega_1 \approx \omega_{1f}$ for this case. In all cases, the Reynolds number remains small where $0.1 \le Re_\omega  \le 4$ which indicates the importance of viscosity as expected for small beams in liquid.

The bottom section of Table~\ref{fit_parameters} includes the experimentally measured, and theoretically predicted, values for the effective mass $m_1$, natural frequency $\omega_1$, and the effective spring constant $k_1$ of the fundamental mode for the different beams. The experimental values of $m_1, \omega_1$, and $k_1$ are directly measured in experiment using air as the surrounding fluid.  The frequency of the fundamental peak determines $\omega_1$ and the equipartition of energy is used to determine an experimental value of $k_1$. Using these two measured value of $\omega_1$ and $k_1$, the experimental measurement of the effective mass is determined from $m_1 = k_1/\omega_1^2$.

The theoretical values of $\alpha_1$, $m_1'$, $\omega_1'$, and $k_1'$ are found using the theory of an Euler-Bernoulli beam with tension that is described in Section~\ref{Beam Equation and Eigenmodes}. We use the  primed variables,  $m_1'$, $\omega_1'$, and $k_1'$, to clearly distinguish between these theoretical predictions and the experimentally determined values, $m_1$, $\omega_1$, and $k_1$. The theoretical value of $\alpha_1$ is found using $\phi_1(1/2)$ in Eq.~(\ref{eq:moverm}). A useful limiting case is that of a string under tension which yields $\alpha_1 = 1/2$. It is clear from our results that as the beam gets longer, the value of $U$ increases, and therefore the value of $\alpha_1$ approaches a value of 1/2. Once $\alpha_1$ is determined, it is straight forward to compute the theoretical values $m_1'$, $\omega_1'$, and $k_1'$.  The theoretical prediction for the effective mass is $m_1' = \alpha_1 m_b\rho_s l b h$. The theoretical prediction for the fundamental frequency is given by Eq.~(\ref{eq:Omega_n}). Lastly, the prediction for the effective spring constant is $k_1'=\alpha_1 \rho_s l b h {\omega_1'}^{2}$.

It can be seen that there is some disagreement between the experimental and theoretical values of the fundamental frequency and a larger disagreement in the values of the effective mass and spring constant. The source of this disagreement depends upon several factors.  An important contribution is our choice of the averaged material properties that we use to describe all of the beams. For example, it is possible to have lower errors in $m_1'$, $\omega_1'$, and $k_1'$ if we were instead to fit for the parameters for each individual beam while using an adjustable parameter, such as an effective length for the beam, and also allowing for Young's modulus to have a value beyond what is expected for a silicon nitride beam.  However, this complicates the description by creating parameter values for each beam and we have not followed this approach here.  We note that $m_1'$, $\omega_1'$, and $k_1'$ are not used in generating the theoretical curves shown in Fig.~(2)-(3) of the main text, and they are computed here only to provide some insight into the ability of an Euler-Bernoulli beam with tension to capture the dynamics of our experimental beams. In addition, there are several factors that we have not included in our analysis. These include the stiffness and the mass of the metallic electrodes as well as the presence of fabrication imperfections such as undercuts.

The theoretical predictions of the susceptibility squared $|\hat \chi(\omega)|^2$ and the PSD of the Brownian fluctuations $G_{W}(\omega)$ presented in Fig.~(2) of the main text require as input $m_1$, $\omega_1$, $k_1$, and $T_0$.  Here, we have used the values for $m_1$, $\omega_1$, and $k_1$ that were determined from experiments conducted in air for each individual beam. The mass loading parameter $T_0$ depends upon the beam width and thickness, the density of the beam, and the density of the fluid.  Of all of these quantities, we are most uncertain of the density of the beam $\rho_s$. We have used the value of $\rho_s$ that has been determined using the approach described in Section~\ref{section:Determining Material Properties from Natural Frequencies}.  

There are several additional possible sources of disagreement between the experimental and theoretical curves of $|\hat \chi(\omega)|^2$ and $G_{W}(\omega)$ near the fundamental mode of oscillation that we would like to highlight. The first is the interaction of the oscillating viscous boundary layer and surrounding solid boundaries~\cite{Clark:2008,Lissandrello2012}. This effect is often called squeeze film damping. This additional source of damping causes a reduction in the quality factor of the oscillator. The effect of squeeze film damping can be quantified using the oscillating boundary layer thickness $\delta_s$ that is generated by the oscillating beam (cf.~\cite{green:2005,clarke:2006,Clark:2008}). The boundary layer thickness scales as  $\delta_s \propto \sqrt{\nu_f/\omega}$, which captures the dependence of  $\delta_s$ upon the kinematic viscosity $\nu_f$ and the frequency of oscillation $\omega$. An estimate shows that $\delta_s \sim 1~\mu \rm m$ at 100 kHz, which is of the same order as the gap distance between the beam and the closest solid surface. An important aspect is that $\delta_s$ varies as the inverse square root of  the frequency. Therefore, the oscillations of the fundamental mode will have a larger boundary layer thickness than the higher modes, and in addition, the longer beams will have the largest boundary layers.  For example, the boundary layer thickness decreases to $\delta_s \sim 300 ~\rm nm$ at 1 MHz. In terms of our results, the presence of this additional damping would cause the peak region near the fundamental mode for curves of $|\hat \chi(\omega)|^2$ and $G_{W}(\omega)$ to widen. This effect should be most prominent for our longest beam with $l=60$ $\mu$m. From Fig. 2(a) of the main text it is evident that the experimental curves are wider than the theory curves which would be expected due to the additional squeeze film damping in the experiment that is not included in the theory.

Another source of error is in the mode shapes of the experimental beams. The experimentally measured mode shapes are distorted when compared with the theoretically predicted mode shapes. This is evident for the fundamental mode as shown in Fig.~\ref{fig:Supp_drive}(b)-(c).  Figure~\ref{fig:Supp_drive}(b) shows the 50 $\mu$m beam that has been driven at both ends of the beam. The mode shape is significantly wider than what is predicted theoretically. Figure~\ref{fig:Supp_drive}(c) shows the experimental mode shape of the 20 $\mu$m beam which has been driven only at one end of the beam near $x=0$. In this case, the fundamental mode shape is skewed toward the edge of the beam near the electrothermal drive. In this case, it can be seen  that when a point measurement is made at the center, the measured amplitude will be $\sim 20\%$ smaller than the predicted one due to the distortion, resulting in a less accurate linear response measurement. 
\begin{table}
\caption{\label{fit_parameters} Parameters used for the calculation of the theoretical $G_{W} (\omega)$ and $\left| {\hat\chi (\omega ) } \right|^2$.}
\begin{ruledtabular}
\begin{tabular}{lcccccccc}
Parameter & Symbol or Formula & Units&  60 $\mu \rm m$ & 50 $\mu \rm m$ & 40 $\mu \rm m$ & 30 $\mu \rm m$ & 20 $\mu \rm m$ & 15 $\mu \rm m$ \\
\hline
Young's Modulus & $E$ & GPa  & 300 & 300 & 300 & 300 & 300 & 300\\
Beam Density & $\rho_s$ & $\rm kg/m^3$ & 2750 & 2750 & 2750 & 2750 & 2750 & 2750\\
Axial Load & $S$ & $\rm \mu N$ & 7.64 & 7.64 & 7.64  & 7.64  & 7.64  & 7.64 \\ 
Temperature  & $T$ & K & 295.15 & 295.15 & 295.15 & 295.15 & 295.15 & 295.15\\
Air Kinematic Viscosity &  $\nu_f$ &  $\rm m^2/s \times 10^{-7}$ & 156 & 156 & 156 & 156 & 156 & 156\\
Water Kinematic Viscosity &  $\nu_f$ &  $\rm m^2/s \times 10^{-7}$ & 9.55 & 9.55 & 9.55 & 9.55 & 9.55 & 9.55\\ 
IPA Kinematic Viscosity  &  $\nu_f$ & $\rm m^2/s \times 10^{-7}$ & - & - & 27.8 & - & - & -\\
Air Density & $\rho_f$ & $\rm kg/m^3$  & - & - & 1.18 & - & - & - \\
Water Density & $\rho_f$ & $\rm kg/m^3$ & 997.8 & 997.8 & 997.8 & 997.8 & 997.8 & 997.8\\
IPA Density & $\rho_f$ & $\rm kg/m^3$  & - & - & 786 & - & - & - \\
\hline
Mass Loading Parameter  in Water & $T_0$ & - & 2.91 & 2.91 & 2.91 & 2.91 & 2.91 & 2.91\\
Mass Loading Parameter in IPA & $T_0$ & - & -& - & 2.29 & - & - & - \\
Tension Parameter & $U$ & - & 720 & 500 & 320 & 180 & 80 & 45 \\
Reynolds Number in Air & ${\rm Re}_{\omega}(\omega_{1})$ & - & 0.13 & 0.16 & 0.21 & 0.32 & 0.49 & 0.75 \\
Reynolds Number in Water & ${\rm Re}_{\omega}(\omega_{1f})$ & - & 0.34 & 0.55 & 0.68 & 1.36 & 2.52 & 3.78  \\
Reynolds Number in IPA & ${\rm Re}_{\omega}(\omega_{1f})$ & - & - & - & 0.11 & - & - & -  \\
\hline
Effective Mass (experiment)  & $m_1$ & pg & 10.82 & 9.63 & 6.53 & 3.33 & 1.83 & 1.26 \\
Fundamental Frequency (experiment) & $\frac{\omega_1}{2\pi}$ & MHz & 1.48 & 1.80 & 2.35 & 3.55 & 5.38 & 8.28 \\
Spring Constant (experiment) & $k_1$ & $\rm N/m$ & 0.93 & 1.24 & 1.42 & 1.66 & 2.09 & 3.42\\                
Effective Mass Coefficient (theory) & $\alpha_1$ & - & 0.476 & 0.469 & 0.462 & 0.451 & 0.434 & 0.423 \\
Effective Mass (theory) &  $m_1'$ & pg & 6.93 & 5.70 & 4.49 & 3.29 & 2.11 & 1.54\\
Fundamental Frequency (theory) & $\frac{\omega_1'}{2\pi}$ & MHz & 1.56  & 1.90 & 2.42 & 3.35 & 5.42 & 7.85 \\
Spring Constant (theory) &  $k_1'$ & N/m & 0.66 & 0.81 & 1.04 & 1.46 & 2.45 & 3.75\\

\end{tabular}
\end{ruledtabular}
\end{table}

\clearpage

\section{Appendix: Additional Data}

In this appendix, we present additional data plots using different axes and non-dimensionalizations.  All of the relevant information is included in the figure captions. 

\begin{figure}[!htb]
   \begin{center}
   \includegraphics[width=6.75in]{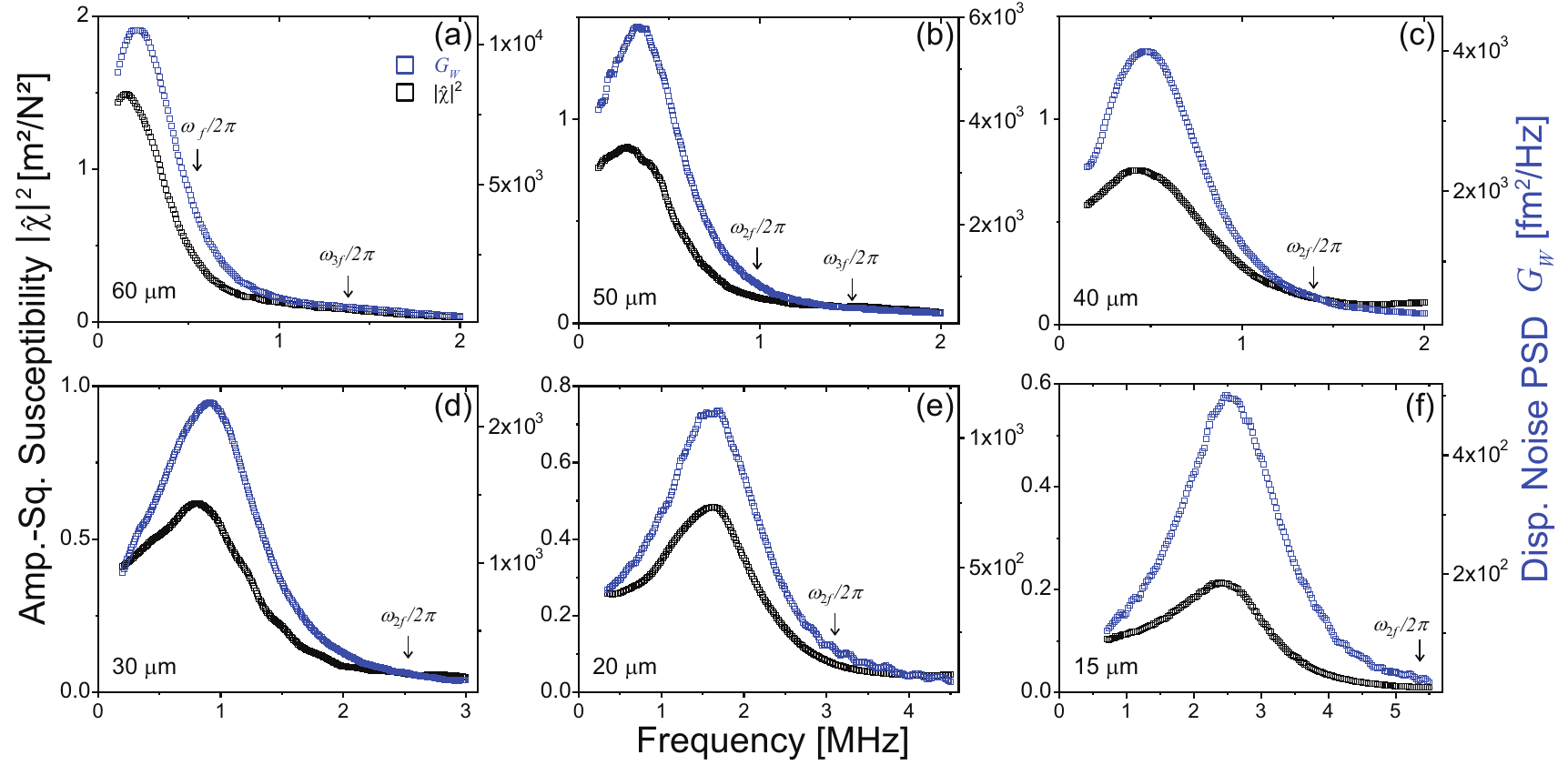} 
    \caption{Figure~2 from main text plotted in linear scale. Amplitude-squared susceptibility $\left|{\hat \chi} \right|^2$ (black) and PSD of the displacement fluctuations $G_{W}$ (blue) for each beam in water; beam length is indicated in the lower left of each sub-figure. Both quantities are measured at the center of the beam. The arrows show the approximate positions of the second ($\omega_{2f}/2\pi$) and third mode ($\omega_{3f}/2\pi$) peaks in fluid (when in range).}
    \end{center}
\end{figure}

\begin{figure}
   \begin{center}
   \includegraphics[width=6.75in]{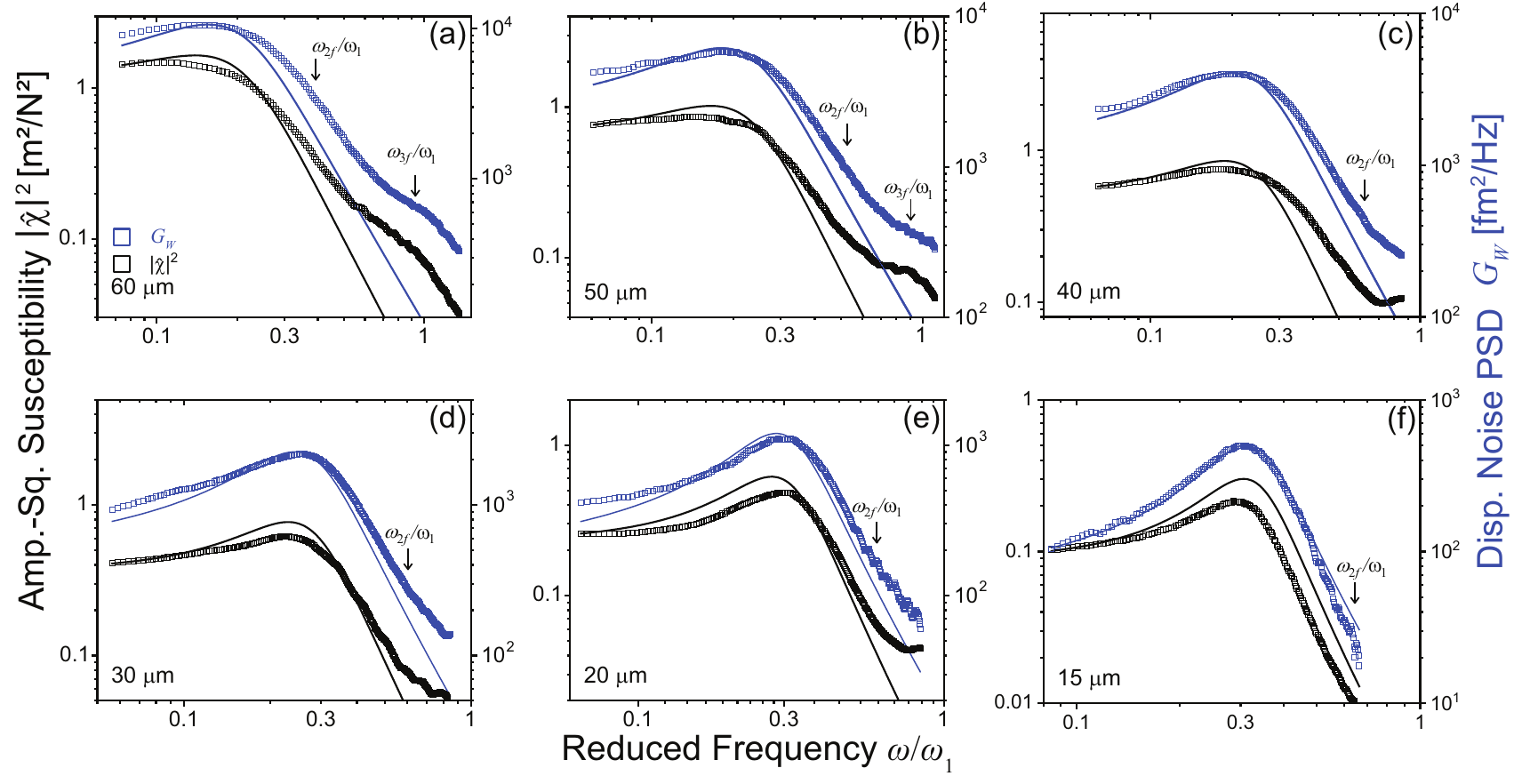} 
    \caption{Figure~2 from main text plotted as a function of  reduced frequency, $\omega / \omega_1$, where $\omega_1$ is the fundamental frequency in air. The arrows show the approximate positions of the second ($\omega_{2f}/\omega_1$) and third mode ($\omega_{3f}/\omega_1$) peaks in fluid (when in range).}
    \end{center}
\end{figure}

\begin{figure}
   \begin{center}
   \includegraphics[width=6.75in]{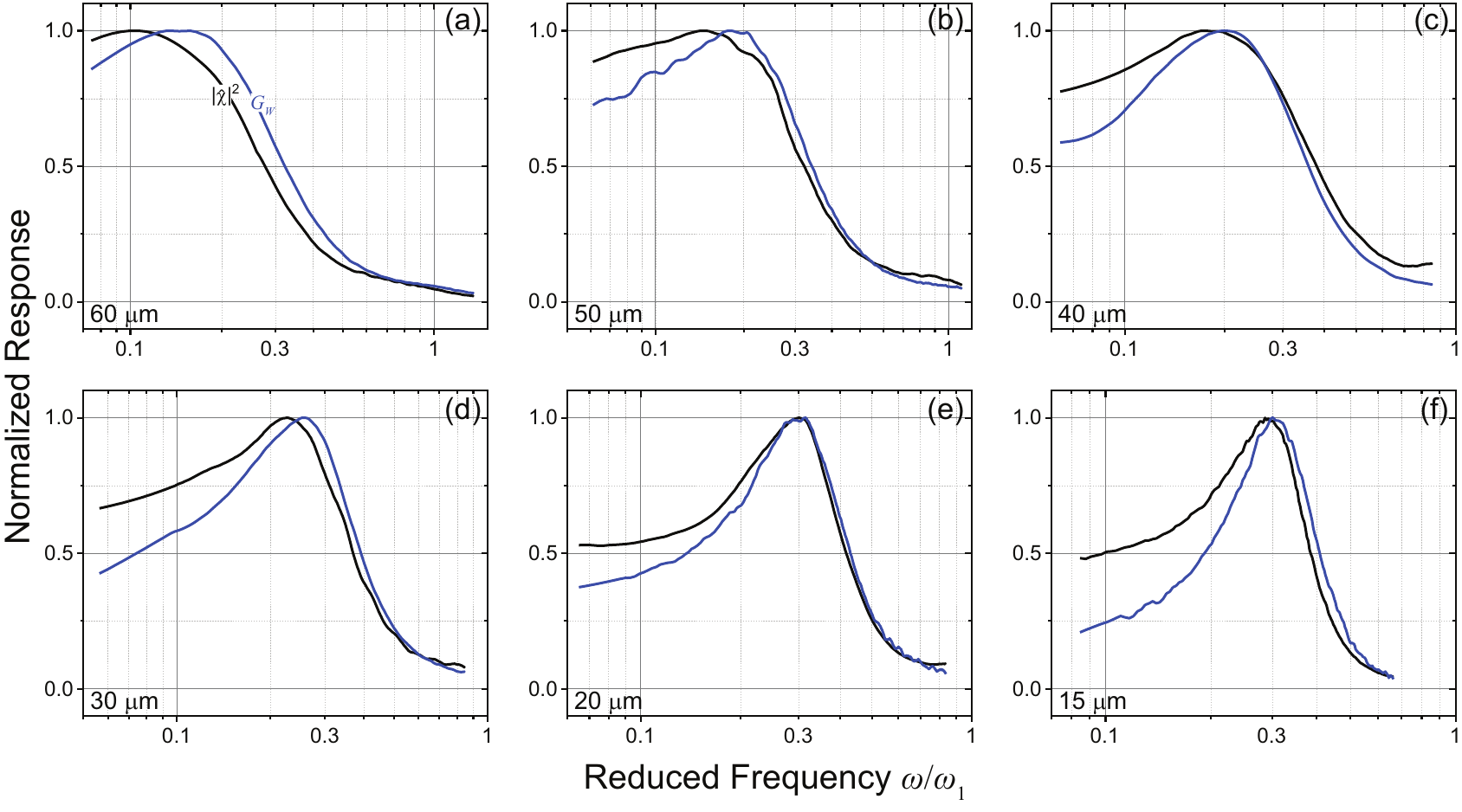} 
    \caption{Normalized amplitude-squared susceptibility $\left|{\hat \chi} \right|^2$ (black)  and PSD of the displacement fluctuations $G_{W}$ (blue) as a function of reduced frequency, $\omega / \omega_1$, for all the devices in water.}
    \end{center}
\end{figure}

\begin{figure}
   \begin{center}
   \includegraphics[width=6.75in]{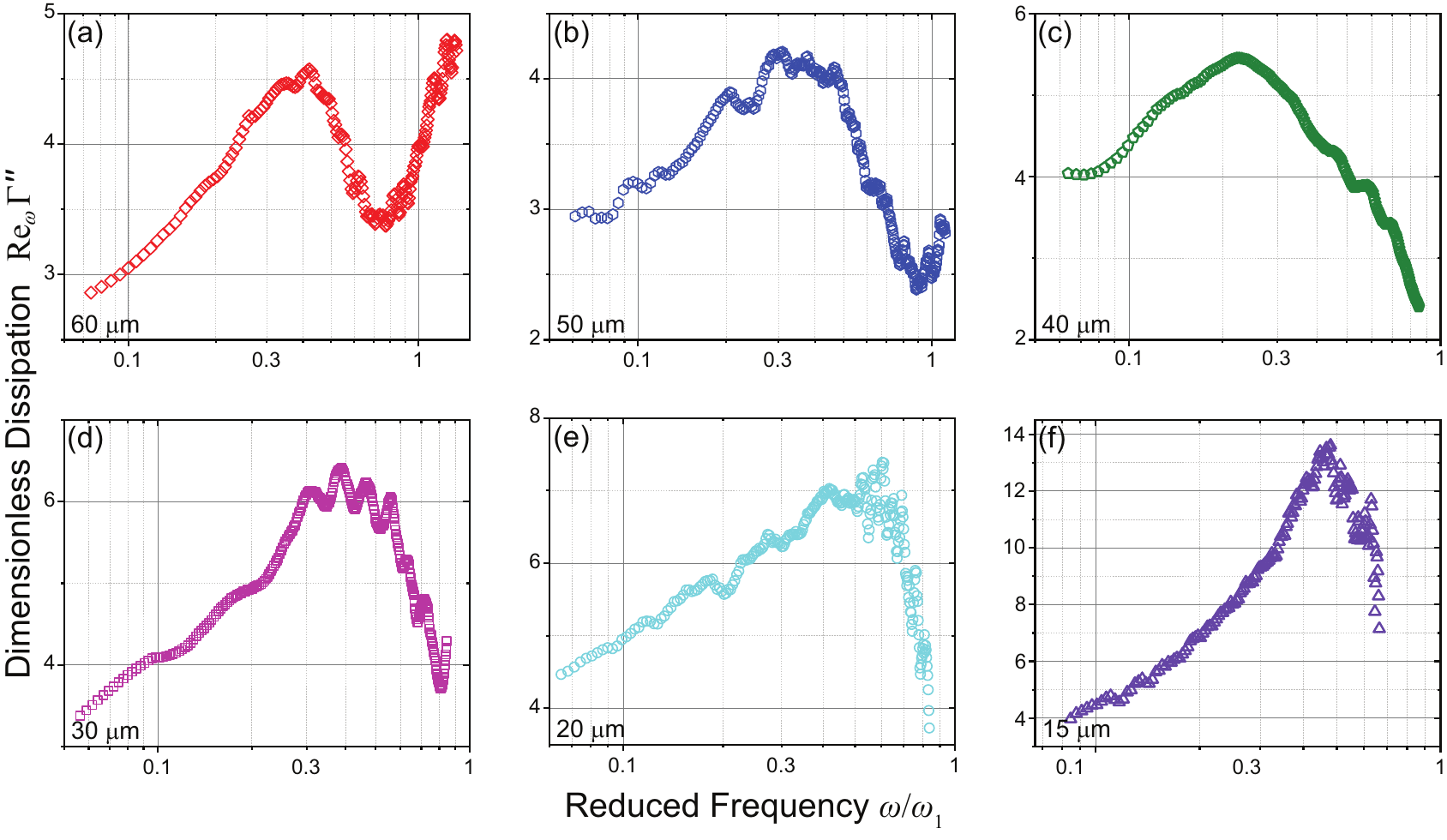} 
    \caption{Semi-logarithmic plots showing the measured dimensionless dissipation for all the devices in water as a function of reduced frequency, $\omega / \omega_1$.}
    \label{fig:Supp_A4}
    \end{center}
\end{figure}

\begin{figure}
   \begin{center}
   \includegraphics[width=6.75in]{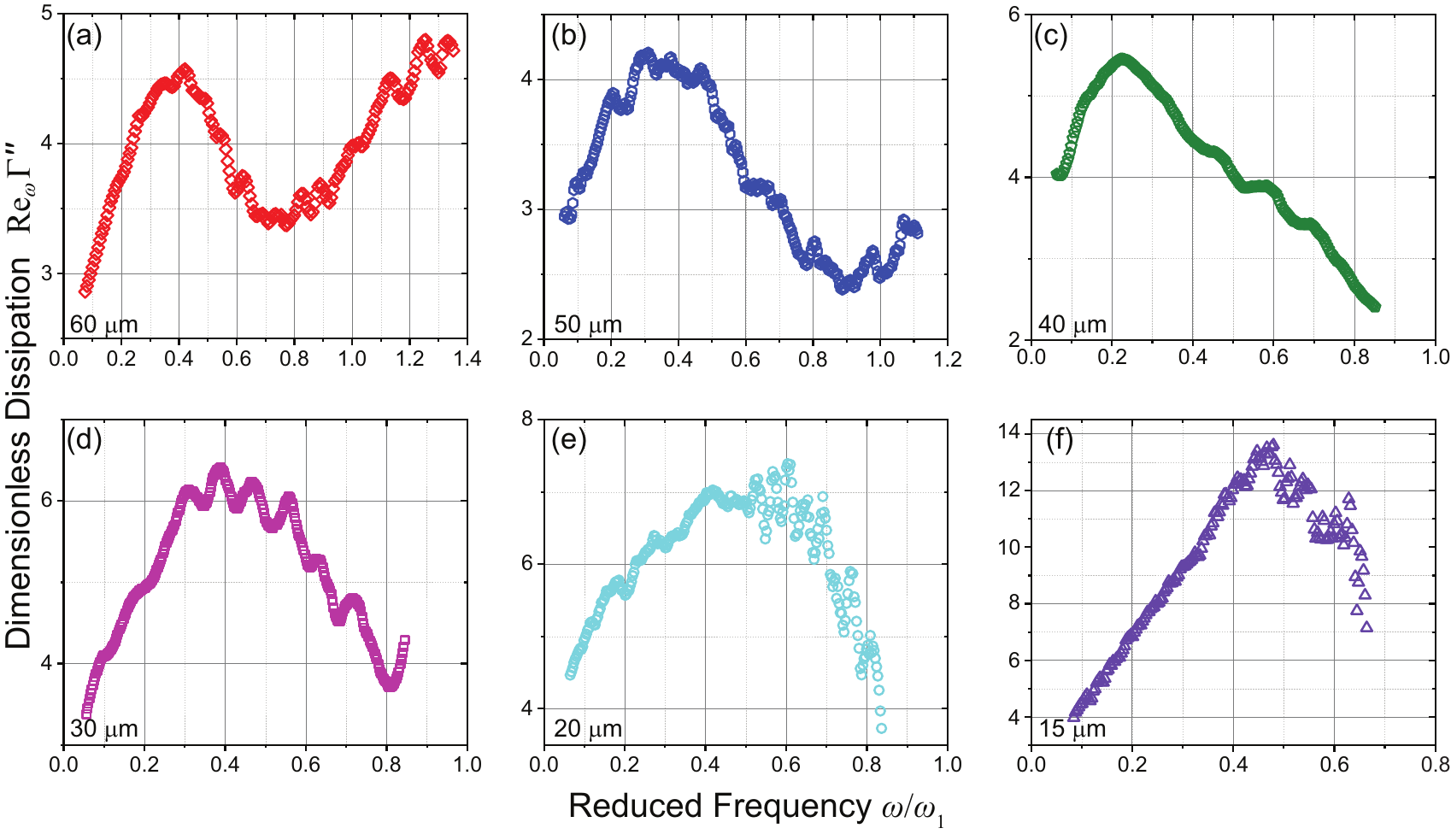} 
    \caption{Linear plots of the  data in Figure~\ref{fig:Supp_A4}.}
    \end{center}
\end{figure}

\clearpage

\bibliography{main.bib}